\documentclass[a4, 11pt, dvipdfm]{article} 
\usepackage{amsmath, amssymb}
\usepackage{abstract}
\usepackage{cite}
\newcommand{\maruk}[1]{\left( #1 \right)}
\newcommand{\mn}[0]{\mu \nu}
\newcommand{\eijk}[0]{\epsilon_{ijk}}
\newcommand{\emn}[0]{\epsilon_{\mu \nu}}
\newcommand{\half}[0]{\frac{1}{2}}
\newcommand{\halfi}[0]{\frac{i}{2}}
\newcommand{\qrt}[0]{\frac{1}{4}}

\makeatletter
    
    \@addtoreset{equation}{section}
 \makeatother
\begin{document}
\thispagestyle{empty}
\begin{flushright}
EPHOU-11-002 \\
April, 2011
\end{flushright}
\vspace*{10mm}
\begin{center}
{\LARGE Mass deformation of twisted super Yang-Mills theory \\ \vspace*{2mm} with fuzzy sphere solution}\\
\vspace*{10mm}
{\large Junji Kato$^{\text{a}}$\footnote[1]{jkato@particle.sci.hokudai.ac.jp}, 
Yoshi Kondo$^{\text{a}}$\footnote[2]{proton@particle.sci.hokudai.ac.jp} and 
Akiko Miyake $^{\text{b}}$\footnote[3]{miyake@ippan.kushiro-ct.ac.jp}. }\\
\vspace*{8mm}
\small{\textit{ $^{\text{a}}$ Department of Physics, Hokkaido University  }}\\
\small{\textit{Sapporo 060-0810, Japan}}\\ 
\vspace*{3mm}
\small{\textit{ $^{\text{b}}$  Department of General Education, Kushiro National College of Technology }} \\
\small{\textit{Kushiro 084-0916, Japan}} \\
\vspace*{8mm}
\end{center} 

\begin{abstract}
We investigate mass deformation of twisted superalgebra of $U(N)$ super Yang-Mills (SYM) theories in several models and in several dimensions, 
motivated by the method formulated in \cite{Sug}. 
We show that there are several ways to perform the deformation, if a model possesses four scalar supercharges
except for two dimensional B-model.
We also evaluate classical vacuum solutions of  the potentials of scalar fields for each model. 
We then find that it is always possible to find fuzzy sphere solution in the theories.
\end{abstract}

\tableofcontents

\newpage
\section{Introduction} 
Presently supersymmetry (SUSY) is believed to be one of the most prospective candidates as
the ingredients beyond the standard model and applied to wide variety of models. 
Since our world is not supersymmetric in fact, SUSY should be broken by some non-perturbative mechanisms.
One promising way to deal with such a non-perturbative aspect is to make use of a lattice simulation. 
Indeed for quantitative evaluation, the lattice simulation is widely used at present
for example in search of non-perturbative effects in QCD and SUSY breaking .
In addition, it is also interesting to investigate AdS/CFT correspondence \cite{Mal, AGMO}
from the point of view at lattice or with non-lattice techniques \cite{HNT, AHNT}.


Even though there are the famous doubling problem and the breakdown of the Leibniz rule on the lattice \cite{NN},
some approaches address the formulation of supersymmetric theory on a lattice (See reviews in \cite{KCU, Feo, Kap}). 
Some insist that the model is supersymmetric invariant up to one or two supersymmetry on the lattice.
Full SUSY will, then, be recovered at continuum limit without any fine-tuning \cite{CKKU1, CKKU2, Sug2, Cat}. 
While there is another ambitious approach, called link approach \cite{DKKN, DFKKS},
which claims the model can be exactly full supersymmetric even on the lattice.
Besides it suggests the relationship between the noncommutativity in the link approach and the nonlocality
in a new formulation \cite{DFKKS2}.

For the simulation of supersymmetric theory on the lattice,
there is a technical problem that a flat direction appears as a classical configuration in the potential of scalar fields \cite{Kan}. 
The presence of the flat direction indeed leads to the breakdown of iteration of computation
because of over-counting of field configurations. Therefore it should be resolved by some methods,
such as addition of mass terms to lift up the potential of scalar fields.
In \cite{KaS, HaK} a soft SUSY-breaking mass is introduced to scalar fields in two dimensional models
in order to control the flat direction. As a result, one can obtain the correct continuum limit.

In \cite{Sug}, they introduced the mass deformation method \cite{MNS} to $N=8, D=2$ twisted SYM theory 
to remove the flat direction and consequently showed some steps to obtain $N=4, D=4$ SYM theory 
formulated on the lattice\footnote{In private communication, 
they said they had also applied the same deformation method to $N=(4,4), D=2$ A-model SYM theory.
This paper is not yet published.}. 
In consequence, they succeeded in introducing mass parameter into superalgebra 
by deforming supercharges to keep some of supersymmetries up to $SU(2)_R$ symmetry.
Then, once the theory is deformed to be nilpotent up to $SU(2)_R$ symmetry\footnote{In this setup, 
models come to respect only two SUSYs regardless of the fact that the original model possesses full SUSY, 
say, sixteen supercharges (eight in other models).},
the action receives mass terms for each field. 
In addition, they focused on the fact that there appears fuzzy sphere configuration for the potential minimum of scalar field.
Utilizing the technique, which is well-known in string theoretical point of view \cite{Mey, BMN}, 
they extended the fuzzy sphere configuration to a noncommutative space.
By careful tuning of a lattice spacing, the mass parameter and a parameter of noncommutativity, 
the model eventually leads to noncommutative $N=4, D=4$ SYM theory, 
where two dimensions correspond to original spacetime 
and the rest of two dimensions to non-commutative ones.
It is also possible to formulate $N=8, D=2$ SYM models, which can be obtained 
without embedding on the fuzzy sphere lattice 
and simply with tuning the mass parameter to be zero to keep the potential well.
They also commented on the case for $N=4, D=2$ in the same procedure.
In \cite{Han}, this mass deformation method is also applied to a model combined with other lattice SUSY approach.

This paper is motivated by the mass deformation technique in \cite{Sug}.
We investigated whether it can be applicable to other SYM models 
and possible to find the fuzzy sphere solution for each model. 
Throughout this paper, we focus on topological twisted SYM theories and choose gauge group as $U(N)$.
Indeed the gauge group is taken freely as far as gauge group is large enough to contain $SU(2)$ as a subgroup.
The models investigate in this paper are $N=4, D=4$ SYM, $N=8, D=3$ SYM, $N=4, D=3$ SYM theories.
More precisely we study A-twisted SYM theories with $16$ SUSY in four and three dimensions
and B-twisted SYM theories with $8$ SUSY in three and two dimensions.

In particular, models have to possess enough number of scalar supercharges to proceed with 
the mass deformation so that these are defined in four dimensions or less.
This is due to the fact that the key element of the mass deformation of superalgebra is to use the model 
which conserve at least two scalar supercharges with opposite ghost number. 
Therefore only theories defined in four dimensions or less can be capable of topological twisting and
of possessing appropriate number of twisted scalar supercharges simultaneously.

This paper is organized in the following way: 
Sec. 2 describes a mass deformation method in detail in $N=4, D=4$ SYM model
 and how fuzzy sphere solution arises.
From Sec. 3 to Sec. 5, we apply the mass deformation to each SYM model.
In Sec. 6, we summarize the results and mention some points.

\section{$N=4, D=4$ $U(N)$ SYM} 

Since we study twisted SYM theories as the models which can be formulated on a lattice,
we summarize a topological twisting of SYM theories.
$N=4, D=4$ $U(N)$ SYM theory is obtained 
through dimensional reduction of $N=1, D=10$ $U(N)$ SYM theory.
Through the dimensional reduction 
the isometry decomposes as $SO(9,1) \to SO(3,1) \otimes SO(6)$ 
where $SO(3,1)$ is Lorentz group and $SO(6)$ is the internal symmetry group.
We here change the isometry $SO(3,1)$ into $SO(4)$, 
because we should consider the lattice action in Euclidean spacetime.
Furthermore, Euclidean symmetry and the internal $R$-symmetry are decomposed into subgroups as,
\begin{align}
SO(4)_E &\sim SU(2)_{left} \otimes SU(2)_{right}, \\
SO(6)_R &\to SU(2)_A \otimes SU(2)_B.
\end{align}
We define A-Twist as a diagonal sum of $SU(2)_{left}$ and $SU(2)_A$ in mathematical notation
namely identifying $SU(2)_{left}$ and $SU(2)_A$. 
Resulting symmetries are
\begin{align}
SU(2)'_{left} \otimes SU(2)_{right} \otimes SU(2)_B,
\end{align}
where $SU(2)_{left} \otimes SU(2)_A \to SU(2)'_{left} $.
As we will see later, a subgroup of $SU(2)_B$ symmetry behaves as ghost number symmetry.
We would rename $SU(2)_B$ as $SU(2)_R$ in this section. 
This twisted SYM theory is called A-model or Vafa-Witten theory \cite{VW, Yam, Lab, Kat}.
The theory possesses twisted supersymmetries whose generators are not spinors but scalars,
vectors and tensors. In mass deformation technique we concentrate only on scalar supercharges.

\subsection{Action}
The twisted action can be off-shell invariant for two scalar supercharges
by introducing two kinds of auxiliary fields, $h_\mu$ and $h^+_A$, 
and can be rewritten into supercharge exact form.
The action is,
\begin{align}
\mathcal{S}^{N=4}_{0} = \int d^4x \text{Tr}
&\left( - \mathcal{D}_\mu v \mathcal{D}_\mu v 
- \qrt \mathcal{D}_\mu v_A \mathcal{D}_\mu v_A 
- \mathcal{D}_\mu \phi \mathcal{D}_\mu \overline{\phi} + (F^+_{\mn})^2 \right. \notag \\
& -i \psi_\mu (\mathcal{D}_\mu \lambda - \mathcal{D}_\nu \lambda^+_{\mn}) + \halfi \phi \{ \psi_\mu, \psi_\mu \}
-\halfi \overline{\phi} \{ \lambda, \lambda \} - \frac{i}{8} \overline{\phi} \{ \lambda^+_A, \lambda^+_A \} \notag \\
& -i C_\mu (\mathcal{D}_\mu \chi - \mathcal{D}_\nu \chi^+_{\mn}) - \halfi \overline{\phi} \{ C_\mu, C_\mu \}
+ \halfi \phi \{ \chi, \chi \} + \frac{i}{8} \phi \{ \chi^+_A, \chi^+_A \} \notag \\
& - \frac{i}{4} v \{ \chi^+_A, \lambda^+_A \} + \frac{i}{64} \Gamma^+_{ABC} v^+_A \{ \chi^+_B, \lambda^+_C \} 
- i v^+_{\mn} \{ C_\mu, \psi_\nu \} \notag \\
& + iv \{ C_\mu, \psi_\mu \} - \frac{i}{4} v^+_A \{ \chi^+_A, \lambda \} + \frac{i}{4} v^+_A \{ \chi, \lambda^+_A \} -iv \{ \chi, \lambda \} \notag \\
&  - \frac{1}{4}[ \phi, v^+_A] [\overline{\phi}, v^+_A] - \frac{1}{32} [v^+_A, v^+_B] [v^+_A, v^+_B ] - \qrt [v^+_A, v][v^+_A, v]  \notag \\
&\left. - [\phi, v] [\overline{\phi}, v] + \qrt [\phi, \overline{\phi} ]^2 
-h_\mu h_\mu - \qrt h^+_A h_A^+ \right), \label{act} \\
= \int d^4x \text{Tr} & 
~s \overline{s} \maruk{ \lambda \chi + C_\mu \psi_\mu + \qrt \chi^+_A \lambda^+_A 
+ \frac{i}{96} \Gamma^+_{ABC} v_A^+ [v_B^+, v_C^+] - 2 v^+_A F_A^+ }. \label{n4d4}
\end{align}

Here, we take $\mu, \nu, \cdots =1,2, 3,4$.
$\mathcal{D}_\mu$ is covariant derivative, and is defined as $\mathcal{D}_\mu \equiv \partial_\mu - i [A_\mu, \cdots]$.
Then, the superscript, in $\lambda^+_B$ and $\chi^+_B$, 
denotes antiself-dual two forms and we use the following notation, 
$\phi^+_A = \phi^+_{\mu\nu} = - \half \epsilon_{\mu \nu \rho \sigma} \phi^{+}_{\rho \sigma}$. Also,
we define a self-dual part of the curvature $F^+_{\mu \nu} \equiv \qrt \delta^+_{\mu \nu, \rho \sigma} F_{\rho \sigma}$,
where $ \delta^+_{\mu \nu, \rho \sigma} 
= \delta_{\mu \rho} \delta_{\nu \sigma} - \delta_{\mu \sigma} \delta_{\nu \rho} - \epsilon_{\mu \nu \rho \sigma}$.
Furthermore, $\Gamma^{+ABC}$ is antisymmetric in suffixes $A, B, C$.

In the action, $(\phi, v, \overline{\phi})$ are scalar bosonic fields which transform as triplet under $SU(2)_R$ symmetry and
$A_\mu$ is gauge boson filed which transform as singlet, then
$(\lambda, \chi)$, $(C_\mu, \psi_\mu)$ and $(\lambda^+_B, \chi^+_B) $ transform as doublet fermionic fields under $SU(2)_R$.
Finally, auxiliary fields are singlet under $SU(2)_R$.

This action possesses the following discrete symmetry as, 
\begin{align}
A_\mu & \to A_\mu, &  v &\to v, \notag \\
\phi &\to - \overline{\phi}, & \overline{\phi} &\to - \phi,  & v^+_A &\to -v^+_A, \notag \\
\chi &\leftrightarrow \lambda, & \chi^+_A &\leftrightarrow \lambda^+_A, & C_\mu &\leftrightarrow \psi_\mu. 
\end{align}
This is the reflection of $SU(2)_R$ symmetry and the presence of $s$ and $\overline{s}$ charges.

Classical configuration of this model is trivial solution for fermionic fields, $v^+_A$ and gauge fields, but
the model possesses flat direction for scalar fields $\phi, v$ and $\overline{\phi}$
due to the potential terms which appear in the last line in (\ref{act}).
\begin{align}
\phi &= v = \overline{\phi} \propto \textbf{1}, & (\text{other fields}) &= 0,
\end{align}
where $\textbf{1}$ is $N \times N$ identity matrix.

We define twisted supercharges as $s$ and $\overline{s}$ and they satisfy the following subalgebra 
up to gauge transformation at off-shell.
\begin{align}
s^2 &=  \overline{s}^2 = \delta_g, & \{s, \overline{s} \} &= \delta_g.
\end{align}
Supercharges $s$ and $\overline{s}$ possess ghost number $+1$ and $-1$ respectively. 
Therefore $s$ and $\overline{s}$ also transform as doublet under $SU(2)_R$.
By integratiing out the auxiliary fields,
twisted $N=4$ supersymmetry with sixteen supercharges are recovered. 
$s$ and $\overline{s}$ transformations of each field are shown in the table in Appendix B.1.

\subsection{Mass deformation of superalgebra}
Next, we introduce deformation terms into supersymmetric charges as,
\begin{align}
Q &= s + \Delta s & \overline{Q}, &= \overline{s} + \Delta \overline{s},
\end{align}
and the deformation terms $\Delta s$ and $\Delta \overline{s}$ satisfy the following algebras,
\begin{align}
(\Delta s)^2 &= (\Delta \overline{s})^2 =0, & \{ \Delta s, \Delta \overline{s} \} &=0. 
\end{align}
We then define the following nontrivial mass deformed transformation 
for auxiliary fields and fermionic fields.
These fermionic fields with nontrivial transformation are selected from the ones 
which do not appear in the transformation of auxiliary fields.
\begin{align}
\Delta s ( h_\mu) &=  M C_\mu, & \Delta \overline{s} (h_\mu) &= M \psi_\mu, \label{msn401} \\
\Delta s (h^+_A) &=  - M \lambda^+_A, & \Delta \overline{s} (h^+_A) &= M \chi^+_A, \\
\Delta s (\lambda) &=  -2 M \phi, & \Delta \overline{s} (\lambda) &= 2M v, \\
\Delta s (\chi) &= -2 M v, & \Delta \overline{s} (\chi) &= -2M \overline{\phi} \label{msn402}, 
\end{align}
where $M$ is a mass parameter.
By adding deformed terms, new supercharges, $Q$ and $\overline{Q}$, satisfy the following algebra
involving $SU(2)$ generators.
\begin{align}
Q^2 &= M J_{++}, & \overline{Q}^2 &= -M J_{--}, & \{ Q, \overline{Q} \} &= - M J_0 = -2 M J_0' \label{comu},
\end{align}
where $J_{++}$ and $J_{--}$ are ladder operators, 
and eigenvalue of $J_0$ corresponds to the ghost number of each field.
Transformations by $J_{++}, J_{--}$ and $J_0$ are shown in Table 1.
As seen in Table 1, fermionic fields, 
($C_\nu, \psi_\nu$), ($\lambda^+_B, \chi^+_B$ ) and ($\lambda, \chi$), form doublet 
and bosons ($\phi, v, \overline{\phi}$) form triplet.

Each coefficient appearing on the right hand side of (\ref{msn401})-(\ref{msn402}) is determined to keep the algebra (\ref{comu}).
A relation corresponding to (\ref{comu}) frequently appear in various models in this paper.
We would therefore keep the same form of the relation (\ref{comu}) in the following sections,
so that the general form of resulting action and other algebra do not receive any changes
by the dimensional reduction.

\begin{table}[h]
\begin{center}
\begin{tabular}{|c|c||c|c|c|}
\hline
 & gh\# & $J_{++}$ & $J_{--}$ & $J_0$ \\ \hline
 $C_\nu$         & $1$   & $0$                & $\psi_\nu$       & $C_\nu$ \\
 $\psi_\nu$      & $-1$ & $C_\nu$          & $0$                 & $-\psi_\nu$ \\ \hline
 $\lambda^+_B$ & $1$   & $0$                & $-\chi^+_B$ & $\lambda^+_B$ \\ 
 $\chi^+_B$      & $-1$ & $-\lambda^+_B$ & $0$                & $-\chi^+_B$ \\ \hline
 $\lambda$       & $1$   & $0$                 & $-\chi$       & $\lambda$ \\  
 $\chi$            & $-1$ & $-\lambda$       & $0$                & $-\chi$ \\ \hline
 $\phi$            & $2$   & $0$                 & $-2v$           & $2\phi$ \\
 $v$                & $0$   & $-\phi$           & $\overline{\phi}$ & $0$ \\
 $\overline{\phi}$ & $-2$ & $2v$              & $0$                 & $-2\overline{\phi}$ \\ \hline
 $v^+_B$ & $0$ & $0$ & $0$ & $0$ \\
 $A_\nu$ & $0$ & $0$ & $0$ & $0$ \\ \hline
 $h_\nu$ & $0$ & $0$ & $0$ & $0$\\
 $h^+_B$ & $0$ & $0$ & $0$ & $0$\\ \hline
\end{tabular}
\caption{Transformation by $J_{\pm \pm}$ and $J_0$}
\end{center}
\end{table}

\subsection{Deformed action} 
Since $Q$ and $\overline{Q}$ also form $SU(2)$ doublet, they satisfy following algebra,
\begin{align}
[J_{++}, Q] &= 0, & [J_{--}, \overline{Q} ] &= 0, \\
[J_{++}, \overline{Q}] &= Q, & [J_{--}, Q ] &= \overline{Q}, \\
[J_0, Q ] &= Q, & [J_0, \overline{Q} ] &= - \overline{Q}.
\end{align}
The action can be rewritten with new supercharges  $Q$ and $\overline{Q}$ as, 
\begin{align}
\mathcal{S}^{N=4} &= (Q \overline{Q} -M) \mathcal{F}_{0} \label{fact}  \\
&= \mathcal{S}^{N=4}_0 + (s \Delta \overline{s} + \Delta s \overline{s} + \Delta s \Delta \overline{s} - M) \mathcal{F}_0,
\end{align}
where
\begin{align}
\mathcal{F}_{0} = \int d^4x \text{Tr} \maruk{ \lambda \chi + C_\mu \psi_\mu + \qrt \chi^+_A \lambda^+_A
+ \frac{i}{96} \Gamma^+_{ABC} v_A^+ [v_B^+, v_C^+] - 2 v^+_A F_A^+ }.
\end{align}

The action (\ref{fact}) is invariant under $Q$ and $\overline{Q}$ transformation as long as 
$\mathcal{F}_0$ has $SU(2)_R$ invariance. For example, applying the operator $Q$ to the action,
we obtain $M \overline{Q} J_{++} \mathcal{F}_0$ term from the commutation relation of $J_{++}$ and $\overline{Q}$.
Since this term must vanish, $SU(2)_R$ invariance of $\mathcal{F}_0$ is required.
$SU(2)_R$ invariance of $\mathcal{F}_0$ can be easily confirmed because $SU(2)_R$ doublet fermions are paired together 
and the other terms are composed of $SU(2)_R$ singlet fields, and therefore each term in $\mathcal{F}_0$ possesses zero
ghost number.

Regardless of the mass deformation in superalgebra, two supersymmetries are still conserved. 
Consequently, the action obtains several terms as, 
\begin{align}
\mathcal{S}^{N=4} 
&= \mathcal{S}^{N=4}_0 +  6M i v [\phi, \overline{\phi}] - 4M^2 (v^2 + \phi \overline{\phi}) \notag \\
& ~~~~ +M \maruk{ 2 \lambda \chi - 2 C_\mu \psi_\mu - \half \chi^+_A \lambda^+_A 
- \frac{i}{96} \Gamma^+_{ABC} v^+_A [v^+_B, v^+_C] + 2 v^+_A F^+_A }.
\end{align}
Because of the second term in (\ref{fact}), $\mathcal{F}_0$ itself appears in the action as mass terms of the fermions, 
besides there are new combinations of scalar fields.

The potential term including $v, \phi$ and $\overline{\phi}$ 
which is analogous with the original work \cite{Sug} is
\begin{align}
\mathcal{V}_1  =
-[\phi, v] [\overline{\phi}, v] + \qrt [\phi, \overline{\phi} ]^2 + 6 M i v [\phi, \overline{\phi}] - 4M^2 (v^2 + \phi \overline{\phi}). \label{spot}
\end{align}
In the next subsection, we will focus on this scalar potential and investigate how classical configuration changes.

\subsection{Fuzzy sphere solution} 

In the previous sections, we supposed that $\phi$ and $\overline{\phi}$ are real-valued independent fields, but here 
we would like to consider $\phi$ and $\overline{\phi}$ are hermitian conjugate each other, namely $\phi^\dagger = \overline{\phi}$.
Although there are some ambiguities in interpretation, 
we would leave this problem for later discussion since it indeed does not affect degrees of freedom in this model.
In addition to this reinterpretation, for simplicity,
we replace anti-hermitian fields $\phi, v,$ and $\overline{\phi}$ into hermitian fields,
$\phi \to i \phi, v \to iv$ and $\overline{\phi} \to i \overline{\phi}$.

Subsequently, the scalar potential (\ref{spot}) becomes,
\begin{align}
\mathcal{V}_1  =
-[\phi, v] [\overline{\phi}, v] + \qrt [\phi, \overline{\phi} ]^2 + 6 M v [\phi, \overline{\phi}] + 4M^2 (v^2 + \phi \overline{\phi}). 
\end{align}

This potential is semi-positive definite because it can be written as sum of perfect squares as,
\begin{align}
\mathcal{V}_1 &= \qrt \maruk{ [\phi, \overline{\phi}] + 4 M v }^2 
    + \maruk{[\phi, v] - 2 M \phi} \maruk{ -[\overline{\phi}, v] - 2 M \overline{\phi} } \notag \\
&= \qrt \maruk{ [\phi, \overline{\phi}] + 4 M v }^2 +  \maruk{[\phi, v] - 2 M \phi}  \maruk{[\phi, v] - 2 M \phi}^\dagger \geq 0. \label{pn4d4}
\end{align}

This potential has a nontrivial vacuum configuration due to the deformation.
We then obtain the conditions:
\begin{align}
[\phi, \overline{\phi}] + 4 M v &=0, \\
[\phi, v] - 2 M \phi &=0, \\
-[\overline{\phi}, v] - 2 M \overline{\phi} &= 0,
\end{align}
such that the solution is analogous to Lie algebraic relations as,
\begin{align}
v &= -2 M L_3, \\
\phi &= \pm 2 M L_+, \\
\overline{\phi} &= \pm 2 M L_-,
\end{align}
where $L_3$ and $L_\pm$ are $N$-dimensional representation of $SU(2)$ generators, 
which is subgroup of $U(N)$ gauge group. 
We find fuzzy sphere solution for four dimensional $N=4$ $U(N)$ SYM theory,
and the flat direction is resolved consequently.

\subsection{Adding $\Delta \mathcal{F}$ terms} 
In mass deformation process, the action has further possibility to add extra terms 
$\Delta \mathcal{F}$ as long as it is $SU(2)_R$ invariant.
\begin{align}
\mathcal{S} &= (Q \overline{Q} - M) (\mathcal{F}_0 + \Delta \mathcal{F}), 
\end{align}
where
\begin{align}
\mathcal{F}_0 &= \int d^4x \text{Tr} \maruk{ \lambda \chi + C_\mu \psi_\mu + \qrt \chi^+_A \lambda^+_A 
+ \frac{i}{96} \Gamma^+_{ABC} v_A^+ [v_B^+, v_C^+] - 2 v^+_A F_A^+ }, \\
\Delta \mathcal{F} &= \tilde{m} (v^+_B v^+_B).
\end{align}
In this particular model, the previous deformation does not produce mass term of $v^+_B$,
so it is reasonable to include mass term of $v^+_B$, 
where $\tilde{m}$ is some arbitrary parameters with mass dimension one.

The action becomes, 
\begin{align}
\mathcal{S}
&= \mathcal{S}^{N=4}_{0} + \maruk{ 6M i v [\phi, \overline{\phi}] - (2M)^2 (v^2 + \phi \overline{\phi}) } \notag \\
& ~~~~ +M \maruk{ 2 \lambda \chi - 2 C^\mu \psi_\mu - \half \chi^+_A \lambda^+_A 
- \frac{i}{96} \Gamma^{+ABC} v^+_A [v^+_B, v^+_C] + 2 v^+_A F^+_A } \notag \\
& ~~~~ + \tilde{m} \maruk{ \frac{i}{32} \Gamma^+_{BCD} v^+_B [v^+_C, v^+_D] 
- i v^+_B [v^+_B, v] -2 v^+_B F^+_B + v^+_B h^+_B }
+ \half \tilde{m} \chi^+_B \lambda^+_B - \tilde{m} M v^+_B v^+_B.
\end{align}
The third line includes new terms due to $\Delta \mathcal{F}$.

After eliminating auxiliary fields by solving equation of motion, we can also focus on the potential terms 
with respect to $v^+_B$.
\begin{align}
\mathcal{V}_2 &= - \frac{1}{32} [v^+_A, v^+_B][v^+_A, v^+_B] + \maruk{ \frac{i}{32} \tilde{m} 
- \frac{i}{96} M }\Gamma^{+}_{ABC} v^+_A [v^+_B, v^+_C]
+ \maruk{ \tilde{m}^2 - \tilde{m} M }v^+_B v^+_B.
\end{align}
By inserting appropriate parameter $\tilde{m} = -\frac{M}{3}$, the action also becomes semi-positive definite as follows.
\begin{align}
\mathcal{V}_2 &= \frac{1}{32} 
\maruk{ [v^+_A, v^+_B] + \frac{i}{3} M \Gamma^{+}_{ABC} v^+_C}^\dagger \maruk{[v^+_A, v^+_B] + \frac{i}{3} M \Gamma^{+}_{ABC} v^+_C},
\end{align}
where a formula, $\Gamma^{+}_{ABC} \Gamma^+_{ABD} = 2\cdot 4^2 \delta^+_{C,D}$, is used.

Since $v^+_B$ is self-dual fields, all the $v^+_B$ are rewritten into $v^+_{i4}$ 
through the relation $v^+_B = v^+_{\mn} = - \half \epsilon_{\mn \rho \sigma} v^{+}_{\rho \sigma}$.
In the end, we can find another fuzzy sphere solution with the formula $\Gamma^+_{i4,j4,k4} = -2 \eijk$,
\begin{align}
v^+_{i4} = \frac{2M}{3} L_i
\end{align}
where $L_i$ is $N$-dimensional representation of $SU(2)$.

Before ending this section it should be noted that choices of fuzzy sphere solution leads to 
a problem concerning to positivity of kinetic terms.
In section 2.4, we redefined some fields from anti-hermitian into hermitian ones.
Since one fuzzy sphere solution is only related to scalar fields $v, \phi$ and $\overline{\phi}$ and 
the other solution is only to the fields $v^+_A$, we independently select hermiticity of these fields to 
obtain the fuzzy sphere solution. However the kinetic term of $v^+_A$ is not positive definite.
Indeed the negative sign of kinetic term causes the instability of the action itself, and therefore
it would be not appropriate to adopt two solutions simultaneously.

\section{$N=8, D=3$ $U(N)$ SYM} 
By dimensional reduction of $N=4, D=4$ SYM theory, new two scalar supercharges, $s_4$ and $\overline{s}_4$, appear
so that it is possible to consider new pairs of supercharges to perform
mass deformation of superalgebra as we have shown in $N=4, D=4$ action.

To analyze isometry of this model, we firstly consider naive dimensional reduction of ten-dimensional $N=1$ SYM theory
down to $D=3$. Global symmetry is,
\begin{align}
SU(2)_E \otimes SO (7).
\end{align}
This internal symmetry can be further decomposed as,
\begin{align}
SO (7) \to SU(2)_R \otimes SU(2)_1 \otimes SU(2)_2.
\end{align}
To obtain desired twisted model, we define a twist as a diagonal sum of $SU(2)_E$ and $SU(2)_2$, 
so that the model possesses isometry of,
\begin{align}
SU(2)'_{E} \otimes SU(2)_R \otimes SU(2)_1,
\end{align}
where $SU(2)_{E} \otimes SU(2)_2 \to SU(2)'_{E}$.
This internal symmetry contains one extra $SU(2)$ symmetry compared to $N=4, D=4$ SYM model.
It suggests that there would be other topological supercharges which correspond to this $SU(2)$ symmetry, 
indeed it corresponds to two scalar supercharges, $s_4$ and $\overline{s}_4$, as noted in the beginning of this section.
In this model, we expect $U(1)$ subgroup of either $SU(2)_R$ or $SU(2)_1$ corresponds to ghost number symmetry.
 
\subsection{Action}  
Three dimensional $N=8$ twisted $U(N)$ SYM action is,
\begin{align}
\mathcal{S}_0^{N=8} = \int d^3x \text{Tr}
& \left[ \frac{\mbox{}}{\mbox{}} - \mathcal{D}_i v \mathcal{D}_i v   
-  \mathcal{D}_i v_{j} \mathcal{D}_i v_{j} + [A_4, v_{j}] [A_4, v_{j} ]  \right. \notag \\
&  - \mathcal{D}_i \phi \mathcal{D}_i \overline{\phi} 
+ \mathcal{D}_i A_4 \mathcal{D}_i A_4 + \half F_{ij} F_{ij} \notag   \\
& -i \psi_i \maruk{ \mathcal{D}_i \lambda + \eijk \mathcal{D}_j \lambda_{k} + i[A_4, \lambda_{i}] }
- i \psi_4 \maruk{ \mathcal{D}_i \lambda_{i} - i [A_4, \lambda] } \notag \\
& -i C_i \maruk{ \mathcal{D}_i \chi + \eijk \mathcal{D}_j \chi_{k} + i[A_4, \chi_{i}] }
-i C_4 \maruk{ \mathcal{D}_i \chi_{i} - i [A_4, \chi] } \notag \\
&  + \halfi \phi \{ \psi_i, \psi_i \} + \halfi \phi \{ \psi_4, \psi_4 \} - \halfi \overline{\phi} \{ \lambda, \lambda \}
- \halfi \overline{\phi} \{ \lambda_{i}, \lambda_{i} \} \notag \\
& - \halfi \overline{\phi} \{ C_i, C_i \} - \halfi \overline{\phi} \{ C_4, C_4 \}
+ \halfi \phi \{ \chi, \chi \} + \halfi \phi \{ \chi_{i}, \chi_{i} \} \notag \\
& -i v \{ \chi_{i}, \lambda_{i} \} - i \eijk v_{i} \{ \chi_{j}, \lambda_{k} \} 
+ i \eijk v_{i} \{ C_j, \psi_k \} - i v_{i} \{ C_i, \psi_4 \} + i v_{i} \{ C_4, \psi_i \} \notag \\
& + iv \{ C_i, \psi_i \} + i v \{ C_4, \psi_4 \} - i v_{i} \{ \chi_{i}, \lambda \} 
+ iv_{i} \{ \lambda_{i}, \chi \} -i v \{ \chi, \lambda \} \notag \\
& + [A_4, v] [A_4, v] + [A_4, \phi] [A_4, \overline{\phi}] - [\phi, v][\overline{\phi}, v] + \qrt [\phi, \overline{\phi}]^2 \notag \\
& \left.  - [\phi, v_{i}] [\overline{\phi}, v_{i}] -\half [v_{i}, v_{j}][ v_{i}, v_{j} ] - [v_{i}, v][v_{i}, v] - h_i h_i - h_4 h_4 -  h^+_{i4} h^+_{i4} \right] , \label{an8d3}
\end{align}
where Lorentz indices run, $i,j = 1, 2, 3$. 
Most of the situations correspond to $N=4, D=4$ case, while there appear new characters.
New scalar fields appear from the four dimensional vector fields, and new vector fields are
induced by the selfdual two-forms in $N=4, D=4$ action, such as $\lambda^+_A$ and $\chi^+_A$,  
because of the self-dual condition: for example, $\lambda^+_{ij} = - \half \epsilon_{\ij k 4} \lambda^+_{k4} $.

In this models, bosonic scalar fields ($\phi, v, \overline{\phi}, A_4$) possess the ghost number as $(+2, 0, -2, 0)$ respectively. 
Fermionic scalar fields $(\lambda, \chi, C_4, \psi_4)$ possess ghost number as
$(+1, -1, +1, -1)$ respectively, and the ghost number of vector fields 
$(C_i, \psi_i, \lambda_i, \chi_i )$ are assigned as
$(+1, -1, +1, -1)$ respectively.

\subsection{Combinations of supercharges} 
In this subsection, we consider pairings of scalar supercharges.
As in the previous section, the action is SUSY invariant with two scalar supercharges 
even after the mass deformation.
Supertransformations of scalar supercharges are shown in Appendix B.2.
Since ghost number of $(s, \overline{s}, s_4, \overline{s}_4)$ is assigned as $(+1, -1, -1, +1)$ respectively,
it is possible to consider four different ways of choosing two supercharges among them. 
In the following, the action can be written as supercharge exact form with these possible combination of supercharges.
\begin{align}
\mathcal{S}^{N=8}_0
&= s \overline{s} \int d^3x \text{Tr} \maruk{ \lambda \chi  + C_4 \psi_4 + C_i \psi_i + \chi_{i} \lambda_{i}
	- \frac{2i}{3} \eijk v_{i} [v_{j}, v_{k}] - 4 v_{i} \mathcal{D}_i A_4 +2 \eijk v_{i} F_{jk} } \label{n8d301} \\
&= \overline{s}_4 s_4 \int d^3 x \text{Tr} \maruk{ \lambda \chi + C_4 \psi_4 - C_i \psi_i - \chi_i \lambda_i
	- \frac{2i}{3} \eijk v_i [v_j, v_k] + 4 v_i \mathcal{D}_i A_4 + 2 \eijk v_i F_{jk} } \label{n8d302} \\
&= s s_4 \int d^3 x \text{Tr} \maruk{ \psi_4 \lambda + C_4 \chi + \lambda_i \psi_i + C_i \chi_i 
	+ 2i \mathcal{T}_{CS} - 4i v_i \mathcal{D}_i v + 2i \eijk v_i \mathcal{D}_j v_k } \label{n8d303} \\
&= \overline{s}_4 \overline{s} \int d^3 x \text{Tr} \maruk{ \lambda \psi_4 - C_4 \chi + \lambda_i \psi_i + C_i \chi_i
	-2i \mathcal{T}_{CS}  - 4i v_i \mathcal{D}_i v - 2i \eijk v_i \mathcal{D}_j v_k } \label{n8d304}
\end{align}
where, in the last two lines, $\mathcal{T}_{CS}$ denotes Chern-Simons term.
\begin{align}
\mathcal{T}_{CS} &= \eijk \maruk{A_i \partial_j A_k - \frac{i}{3} A_i [A_j, A_k] }.
\end{align}
Thus, we can perform the mass deformation about these actions.
There are some comments about the above combinations:
the first combinations is expected results, since it can be derived straightforwardly 
from dimensional reduction of $N=4, D=4$ model (\ref{n4d4}).  
The second model is composed with $\mu =4$ component of vector supercharges, $s_4$ and $\overline{s}_4$.
On the other hand, the last two are totally new results. 
One remarkable point is Chern-Simons term appears in the integrands,
so this term would play a role to give a mass to the gauge field \cite{Jack} in a deformed action 
in the end.


However, it is not necessary to study all these deformations.
The action (\ref{an8d3}) is invariant under some discrete symmetries,
which may be the consequence of the fact that the action possesses two $SU(2)$ symmetries in internal symmetry.
There are indeed three kinds of discrete symmetries and its details are shown in Appendix A.1. 
With this useful tool, 
we can show equivalence between the deformation of $(s, \overline{s})$ and $(s_4, \overline{s}_4)$
and between $(s, s_4)$ and $(\overline{s}, \overline{s}_4)$
\footnote{One of discrete symmetries makes the model invariant, and other symmetries changes the models to the corresponding partner after the deformation. But these discrete symmetries do not completely show 
that all models are equivalent. 

However, there seems to be nontrivial "\textit{discrete symmetry}" which connects remaining two, for example (\ref{n8d301})  and (\ref{n8d303}). 
This \textit{symmetry} corresponds to taking complex conjugation of complexified covariant derivatives, 
namely $\mathcal{D}^\pm_\mu \equiv \partial_\mu - i (A_\mu \pm i v_\mu)$. 
Interchanging of $\mathcal{D}^\pm_\mu$ results in the action invariant, and it also makes the relation among supercharge as, $s \to s_4 \to \overline{s}_4 \to \overline{s} \to -s$.
This might be connected with $SO(4)$ symmetry since R-symmetry is $SU(2) \otimes SU(2)$, which is isomorphic to $SO(4)$.
But we should carefully examine that there seems to be no reasonable representation of the symmetry just like interchanging of fields, and the origin of the rotational symmetry of supercharges has not been obvious yet. }.

\subsection{Mass deformation of superalgebra in $(s, \overline{s})$ combination} 
We first consider $(s, \overline{s})$ pair 
because the combination of $(s_4, \overline{s}_4)$ is equivalent with it through discrete symmetries.

Let define the following deformation of algebra, 
\begin{align}
Q &= s + \Delta s, & \overline{Q} &= \overline{s} + \Delta \overline{s},
\end{align}
and impose the following algebra in the same manner as the privious section:
\begin{align}
Q^2 &= M J_{++}, & \overline{Q}^2 &= -M J_{--}, & \{ Q, \overline{Q} \} &= - M J_0.
\end{align}
We can define the following nontrivial transformations of deformed terms to satisfy the above algebra.
\begin{align}
\Delta s (h_4) &= M C_4, & \Delta \overline{s} (h_4) &= M \psi_4, \\
\Delta s (h_i) &= M C_i, & \Delta \overline{s} (h_i) &= M \psi_i, \\
\Delta s (h^+_{i4}) &= -M \lambda_i, & \Delta \overline{s} (h^+_{i4}) &= M \chi_i, \\
\Delta s (\lambda) &= -2M \phi, & \Delta \overline{s} (\lambda) &= 2M v, \\
\Delta s (\chi) &= -2 M v, & \Delta \overline{s} (\chi) &= -2M \overline{\phi}. 
\end{align}
This deformed algebra can also be directly derived by the dimensional reduction of $N=4, D=4$ model.
The final result is indeed equivalent to the one which is obtained from dimensional reduction.

The mass deformed action becomes,
\begin{align}
\mathcal{S}_{s, \overline{s}} 
&= (Q \overline{Q} - M) \mathcal{F}_{s, \overline{s}},
\end{align}
where
\begin{align}
\mathcal{F}_{s, \overline{s}} &=\int d^3x \text{Tr} \maruk{ \lambda \chi  + C_4 \psi_4 + C_i \psi_i + \chi_{i} \lambda_{i}
	- \frac{2i}{3} \eijk v_{i} [v_{j}, v_{k}] - 4 v_{i} \mathcal{D}_i A_4 +2 \eijk v_{i} F_{jk}}.
\end{align}
The explicit form of the action is,
\begin{align}
\mathcal{S}_{s, \overline{s}} 
&= \mathcal{S}^{N=8}_0 + 6 i M v  [\phi, \overline{\phi}] - 4 M^2 (v^2 + \phi \overline{\phi}) \notag \\
& ~~~~ -2 M \maruk{ -\lambda \chi  + C_4 \psi_4 + C_i \psi_i + \chi_{i} \lambda_{i}
	- \frac{i}{3} \eijk v_{i} [v_{j}, v_{k}] - 2 v_{i} \mathcal{D}_i A_4 + \eijk v_{i} F_{jk} }.
\end{align}

The potential terms containing scalar fields are,
\begin{align}
\mathcal{V}_{s, \overline{s}} 
&= [A_4, v] [A_4, v] +  [A_4, \phi] [A_4, \overline{\phi}] - [\phi, v][\overline{\phi}, v] 
+ \qrt [\phi, \overline{\phi}]^2 - 4 M^2 (v^2 + \phi \overline{\phi})
 + 6 i M v [\phi, \overline{\phi}] 
\end{align}

In this case, we would focus on $(\phi, v, \overline{\phi})$ which are triplet under $SU(2)_R$.
Similarly we can find the fuzzy sphere solution, 
since in the following potential the last two terms are the same as the potential (\ref{pn4d4}).
\begin{align}
\mathcal{V}_{s, \overline{s}} 
&= -[A_4, v] [A_4, v] - [A_4, \phi] [A_4, \overline{\phi}] 
+ \qrt \maruk{ [\phi, \overline{\phi}] + 4 M v }^2 + \maruk{ [\phi, v] - 2 M \phi} \maruk{ [\phi, v] - 2 M \phi}^\dagger,
\label{vssbar}
\end{align}
where redefinition of the fields $(\phi, v, \overline{\phi})$ are already done as 
in the previous section: $\Phi \to i \Phi, \Phi \in \{ \phi, v, \overline{\phi} \}$.
In this potential, $(\phi, v, \overline{\phi})$ contribute to the fuzzy sphere configuration, 
while classical configuration of $A_4$ is trivial. 

In the following subsections, 
we only consider $(s, s_4)$ combination in the deformation. 
As mentioned above, $(s,\overline{s})$ pair is equivalent to $(s_4, \overline{s}_4)$,
and $(s, s_4)$ pair is also equivalent to $(\overline{s}, \overline{s}_4)$ through the discrete symmetries.
A procedure of the deformation of $(s, s_4)$ combination is the same as the previous one, 
so only the results are shown.

\subsection{Mass deformation of superalgebra in $(s, s_4)$ combination} 
In this subsection, we focus on the cases where Chern-Simons term appear in the integrand.
Similar to the previous procedure, let define the deformed supercharges as,
\begin{align}
Q^+ &= s + \Delta s, & Q^- &= s_4 + \Delta s_4.
\end{align}
We then impose the following algebra, 
\begin{align}
(Q^+)^2 &= M J_{++}, & (Q^-)^2 &= -M J_{--}, & \{ Q^+, Q^- \} &= - M J_0,
\end{align}
so that we define the transformations as,
\begin{align}
\Delta s (L) &= - M \lambda, & \Delta s_4 (L) &= M \psi_4, \\
\Delta s (N_i) &=  M \lambda_i, & \Delta s_4 (N_i) &= M \psi_i, \\
\Delta s (M_i) &=  M C_i, & \Delta s_4 (M_i) &= M \chi_i, \\
\Delta s (C_4) &= -2 M \phi, & \Delta s_4 (C_4) &= 2i M A_4, \\
\Delta s (\chi) &= -2i M A_4, & \Delta s_4 (\chi) &= -2 M \overline{\phi}.
\end{align}
where $N_i, M_i$ and $L$ are auxiliary fields.
In this case, $A_4$ takes part in the transformation instead of $v$.

The deformed action takes the form,
\begin{align}
\mathcal{S}_{s, s_4} 
&= (Q^+ Q^- - M) \mathcal{F}_{s, s_4},
\end{align}
where,
\begin{align}
\mathcal{F}_{s, s_4} &= \int d^3 x \text{Tr} \maruk{ C_4 \chi + \psi_4 \lambda +  \lambda_i \psi_i + C_i \chi_i 
	+ 2i \mathcal{T}_{CS} - 4i v_i \mathcal{D}_i v + 2i \eijk v_i \mathcal{D}_j v_k }, \\
\mathcal{T}_{CS} &= \eijk \maruk{A_i \partial_j A_k - \frac{i}{3} A_i [A_j, A_k] }.
\end{align}
The final result of the action is,
\begin{align}
\mathcal{S}_{s, s_4} 
&= \mathcal{S}^{N=8}_0 - 6M A_4 [\phi, \overline{\phi}] + 4 M^2 (A_4^2 -\phi \overline{\phi}) \notag \\
& ~~~~ -2M \maruk{ -C_4 \chi + \psi_4 \lambda +  \lambda_i \psi_i + C_i \chi_i 
	+ i \mathcal{T}_{CS} - 2i v_i \mathcal{D}_i v + i \eijk v_i \mathcal{D}_j v_k }.
\end{align}
The potential terms containing scalar fields are, 
\begin{align}
\mathcal{V}_{s, s_4} 
&= [A_4, v] [A_4, v] - [\phi, v][\overline{\phi}, v] 
+ \qrt \maruk{ [\phi,\overline{\phi} ] + 4 M A_4}^2 + ([\phi, A_4] - 2 M \phi) ([\phi, A_4] - 2 M \phi)^\dagger.
\end{align}
The fields $\phi$ and $\overline{\phi}$ are already redefined as hermitian. 
In this model, $(\phi, A_4, \overline{\phi})$ transform as triplet under $SU(2)_R$ and contribute to the fuzzy sphere solution.
On the other hand, the classical solution of $v$ is trivial.

The followings are the potential terms including scalar fields in each combination of supercharges:
\begin{align*}
\mathcal{V}_{s, \overline{s}} 
&= [A_4, v] [A_4, v] +  [A_4, \phi] [A_4, \overline{\phi}] - [\phi, v][\overline{\phi}, v] 
+ \qrt [\phi, \overline{\phi}]^2 + 6 M iv [\phi, \overline{\phi}] - 4 M^2 (v^2 + \phi \overline{\phi}),  \\
\mathcal{V}_{\overline{s}_4, s_4} 
&= [A_4, v] [A_4, v] + [A_4, \phi] [A_4, \overline{\phi}] - [\phi, v][\overline{\phi}, v] 
+ \qrt [\phi, \overline{\phi}]^2 -6M iv [\phi, \overline{\phi}] - 4 M^2 (v^2 + \phi \overline{\phi}), \\
\mathcal{V}_{s, s_4} 
&= [A_4, v] [A_4, v]  + [A_4, \phi] [A_4, \overline{\phi}] - [\phi, v][\overline{\phi}, v]
+ \qrt [\phi, \overline{\phi}]^2 - 6M A_4 [\phi, \overline{\phi}] + 4 M^2 (A_4^2 -\phi \overline{\phi}), \\
\mathcal{V}_{\overline{s}_4, \overline{s}} 
&= [A_4, v] [A_4, v]  + [A_4, \phi] [A_4, \overline{\phi}] - [\phi, v][\overline{\phi}, v]
+ \qrt [\phi, \overline{\phi}]^2 - 6M A_4 [\phi, \overline{\phi}] + 4 M^2 (A_4^2 -\phi \overline{\phi}).
\end{align*}
Subscripts of $\mathcal{V}$ denote the combination of the supercharges.
As seen in the list of the potential terms, these potentials lead to the same solution
except that there is an interchange of $v$ and $A_4$. 
There are signs difference in some terms but they are absorbed in a redefinition of the parameter $M$.
It is then concluded that the models considered in this section have the same fuzzy sphere solution consequently.

\section{$N=4, D=3$ $U(N)$ SYM} 
From here mass deformation procedure for three dimensional $N=4$ SYM is treated.
Since $N=2, D=4$ SYM model does not possess two scalar supercharges, 
it is not possible to perform the deformation. We then consider dimensional reduction of
$N=2, D=4$ SYM model down to three dimensions
so that one scalar supercharge arises from vector supercharge.

Three dimensional $N=4$ SYM model is obtained through dimensional reduction 
either directly from $N=1, D=6$ SYM or via $N=2, D=4$ SYM.
This theory is invariant under eight supercharges, namely half SUSY of the previous sections.
And the isometry is given as, from dimensional reduction of $N=1, D=6$ theory,
\begin{align}
SO(6)_E \otimes SU(2)_R 
\to~ &  SO(3)_E  \otimes  SO(3)_N \otimes  SU(2)_R \notag \\
\sim~ & SU(2)_E \otimes  ( SU(2)_N \otimes SU(2)_R)  
\end{align}
where $SU(2)_R$ is R-symmetry in $N=1, D=6$ SYM.

Since there are two $SU(2)$ internal symmetries,
there are two possible twists, in other words diagonal sum of two subgroups, 
choosing either $SU(2)_R$ or $SU(2)_N$ with $SU(2)_E$.
\begin{itemize}
\item A-model (or so called Super-BF \cite{Wit, BBT, BT1}): Twist $SU(2)_E$ with $SU(2)_R$
\begin{align}
SU(2)_E \otimes  SU(2)_N \otimes SU(2)_R \to SU(2)_{E'} \otimes SU(2)_N
\end{align}
Field content consists of $SU(2)_N$ doublet scalar fermions, $SU(2)_N$ doublet vector fermions 
and $SU(2)_N$ triplet scalar bosons with appropriate numbers of auxiliary fields to adjust the degree of freedom.
There are two scalar supercharges which transform as doublets of $SU(2)_N$.
	
\item B-model \cite{BT2}: Twist $SU(2)_E$ with $SU(2)_N$
\begin{align}
SU(2)_E \otimes SU(2)_N \otimes SU(2)_R \to SU(2)_{E'} \otimes SU(2)_R
\end{align}
This model contains of $SU(2)_R$ doublet scalar fermions, $SU(2)_R$ doublet vector fermions
and $SU(2)_R$ singlet vector bosons with appropriate numbers of auxiliary fields.
There are two scalar supercharges which transform as doublets of $SU(2)_R$.
\end{itemize}

In each twisting the model possesses one $SU(2)$ symmetry (either $SU(2)_R$ or $SU(2)_N$) as internal symmetry, 
so that the models commonly possess two scalar supercharges, transforming as doublet under its $SU(2)$ symmetry. 
Therefore, it is expected to perform mass deformation 
and find a fuzzy sphere solution in each model.

\subsection{A-model} 
The action is shown as,
\begin{align}
\mathcal{S}^A_{0} = \int d^3x \text{Tr} 
& \left( \half F_{\mn} F_{\mn} + \mathcal{D}_\mu N \mathcal{D}_\mu N + \mathcal{D}_\mu \phi \mathcal{D}_\mu \overline{\phi}
-i \psi_\mu \mathcal{D}_\mu \psi - i \chi_\mu \mathcal{D}_\mu \chi 
- i \epsilon_{\mu \nu \rho} \psi_\mu \mathcal{D}_\nu \chi_\rho \right. \notag \\
& ~~ -H_\mu H_\mu  +\halfi \overline{\phi} \{ \psi_\mu, \psi_\mu \}
+ \halfi \overline{\phi} \{ \chi , \chi \} + \halfi \phi \{ \chi_\mu, \chi_\mu \} + \halfi \phi \{ \psi, \psi \} \notag \\
& \left. ~~ + N \{ \psi, \chi \} - N \{ \psi_\mu, \chi_\mu \} - [N, \overline{\phi}][N, \phi] + \qrt [\phi, \overline{\phi}]^2 \right) \\
= \int d^3x \text{Tr} &~ s \overline{s}
\maruk{\chi \psi + \psi_\mu \chi_\mu 
+ 2 i \epsilon_{\mu \nu \rho} \maruk{ A_\mu \partial_\nu A_\rho - \frac{i}{3} A_\mu [A_\nu, A_\rho] } } \label{n4d3a},
\end{align}
where $(\phi, N, \overline{\phi})$ are bosonic scalar fields transforming as triplet under $SU(2)_R$, 
besides $A_\mu$ and $H_\mu$ are gauge fields and auxiliary fields respectively, 
both transforming as singlet under $SU(2)_R$.
While $(\chi, \psi)$ and $(\psi_\mu, \chi_\mu)$ are fermionic scalar and vector fields respectively, transforming as
doublet under $SU(2)_R$.

This model possesses two scalar supercharges, out of eight supercharges, $s, \overline{s}, s_\mu$ and $\overline{s}_\mu$.
Similar to the previous section, new supercharges are defined as, 
\begin{align}
Q &= s + \Delta s, & \overline{Q} &= \overline{s} + \Delta \overline{s}.
\end{align}
Supercharges and deformation terms satisfy the following algebra;
\begin{align}
s^2 &= \overline{s}^2 =  \{ s, \overline{s} \} = \delta_g, \notag \\
\Delta s^2 &= \Delta \overline{s}^2=  \{\Delta s, \Delta\overline{s} \} = 0.
\end{align}
In this model, supercharges are nilpotent up to gauge transformation.

Let define the transformation of deformation terms as,
\begin{align}
\Delta s (H_\mu) &= - M \psi_\mu & \Delta \overline{s} (H_\mu) &= - M \chi_\mu, \\
\Delta s (\chi) &= -2 M \phi & \Delta \overline{s} (\chi) &= 2i M N, \\
\Delta s (\psi) &= -2i M N & \Delta \overline{s} (\psi) &= 2 M \overline{\phi}.
\end{align}
 
The action becomes, as far as $Q$ and $\overline{Q}$ satisfy the algebra consistently used in the previous sections,
\begin{align}
\mathcal{S} &= (Q \overline{Q} - M) \mathcal{F}^A_0 \\
&= \mathcal{S}^A_0 + 6M N [\phi, \overline{\phi} ] + 4 M^2 (N^2 + \phi \overline{\phi}) \notag \\
& ~~~~ - 2 M \maruk{ -\chi \psi + \psi_\mu \chi_\mu 
+ i \epsilon_{\mu \nu \rho} \maruk{A_\mu \partial_\nu A_\rho - \frac{i}{3} A_\mu [A_\nu, A_\rho]  } },
\end{align}
where
\begin{align}
\mathcal{F}^A_0 &= \chi \psi + \psi_\mu \chi_\mu 
+ 2 i \epsilon_{\mu \nu \rho} \maruk{ A_\mu \partial_\nu A_\rho - \frac{i}{3} A_\mu [A_\nu, A_\rho] }.
\end{align}

The potential terms including scalar fields are,
\begin{align}
\mathcal{V}_A &= - [N, \overline{\phi}] [N, \phi] + \qrt [\phi, \overline{\phi}]^2 
+ 6 M N [\phi, \overline{\phi} ] + 4 M^2 (N^2 + \phi \overline{\phi}) .
\end{align}
However the notation is different to $N=4, D=4$ model,
this potential is the same as that in $N=4, D=4$ model and $N=8, D=3$ model, 
and the fuzzy sphere solution is also obtained.

\subsection{B-model} 
Next, we consider another twisted model (B-twisted model).
$N=2, D=4$ B-model possesses two scalar supercharges, 
but these supercharges have the same ghost number.
Thus exact part of the action corresponding to $\mathcal{F}_0$ possesses nonzero ghost number.
However $D=3$ B-twisted model possesses two scalar supercharges whose
ghost number is opposite.
Therefore it is possible to perform the mass deformation.

The B-model action is,
\begin{align}
\mathcal{S}^B_0 = \int d^3 x \text{Tr} & 
\left( \qrt F_{\mn} F_{\mn} + \half \mathcal{D}_\mu V_\nu \mathcal{D}_\mu V_\nu 
-2i \lambda_\mu \mathcal{D}_\mu \tilde{\lambda} - 2i \tilde{\lambda}_\mu \mathcal{D}_\mu \lambda \right. \notag \\
& ~~ -2i \epsilon_{\mn \rho} \lambda_\mu \mathcal{D}_\nu \tilde{\lambda}_\sigma + 2i V_\mu \{ \lambda_\mu, \tilde{\lambda} \}
+ 2i V_\mu \{ \tilde{\lambda}_\mu, \lambda \} \notag \\
& ~~ \left. + 2i \epsilon_{\mn \rho} V_\mu \{ \lambda_\nu, \tilde{\lambda}_\rho \} 
- \qrt [V_\mu, V_\nu]^2 + 2K^2 + 2 G \tilde{G} \right), \\
= \int d^3 x \text{Tr} &~ s \overline{s}
\maruk{ -2 \lambda \tilde{\lambda} - 2 \lambda_\mu \tilde{\lambda}_\mu + \epsilon_{\mn \rho} F_{\mn} V_\rho
-i \epsilon_{\mn \rho} \maruk{ A_\mu \partial_\nu A_\rho - \frac{1}{3} A_\mu [A_\nu, A_\rho] } }.
\end{align}
This model includes gauge field $ A_\mu $, bosonic vector field $ V_\mu $,
fermionic fields $ \lambda, \tilde{\lambda}, \lambda_\mu$ and $ \tilde{\lambda}_\mu$ 
and three scalar auxiliary fields $G, \tilde{G}$ and $K$.
These fields are also classified to $SU(2)_N$ doublet scalar fermions, 
$SU(2)_N$ doublet vector fermions, $SU(2)_N$ singlet vector bosons.
What is special in this model is that auxiliary fields possess ghost numbers,
 and they transform as triplet under $SU(2)_N$.

Classical configuration of this model is trivial and there exists flat direction in the solution of
$-\qrt [V_\mu, V_\nu]^2$ term, so that we should apply mass deformation method.
A special feature of this model is that there are vector bosons and no scalar bosonic fields, 
which is slightly different to the previous one, especially in treatment of fuzzy sphere solution. 

Supercharges satisfy the following algebra,
\begin{align}
s^2 = \overline{s}^2 = \{ s, \overline{s} \} &=0.
\end{align}
The action and fields are strictly nilpotent without any gauge transformation, and transformation is shown in Appendix B.3.
Due to internal symmetry of $SU(2)_N$, these scalar supercharges $s$ and $\overline{s}$ possess ghost number $+1$ and $-1$ respectively.

Then, we define the deformed supercharges as,
\begin{align}
Q &= s + \Delta s, & \overline{Q} &= \overline{s} + \Delta \overline{s}, 
\end{align} 
and consider the deformed transformation respecting the following algebra.
\begin{align}
Q^2 &= M J_{++}, & \overline{Q} &= - M J_{--}, & \{ Q, \overline{Q} \} &= - M J_0.
\end{align}
As a consequence, we define the following transformations.
\begin{align}
\Delta s (G) &= 0, & \Delta \overline{s} (G) &= 2 M \lambda, \\
\Delta s (\tilde{G}) &= -2 M \tilde{\lambda}, & \Delta \overline{s} (\tilde{G}) &=0, \\
\Delta s (K) &= M \lambda, & \Delta \overline{s} (K) &= M \tilde{\lambda}, \\
\Delta s (\lambda_\mu) &= 0, & \Delta \overline{s} (\lambda_\mu) &= - M V_\mu, \\
\Delta s (\tilde{\lambda}_\mu) &= M V_\mu, & \Delta \overline{s} (\tilde{\lambda}_\mu) &= 0.
\end{align}
The transformation of auxiliary fields should vanish because auxiliary fields $G$ and $\tilde{G}$
possess ghost number $+2$ and $-2$ and there is no field possessing ghost number $\pm 3$ in this model. 
Even though there are some differences, the mass deformation can be performed. 

The deformed action is, 
\begin{align}
\mathcal{S} &= \maruk{Q \overline{Q} - M} \mathcal{F}_0, 
\end{align}
where,
\begin{align}
\mathcal{F}_0 &= -2 \lambda \tilde{\lambda} - 2 \lambda_\mu \tilde{\lambda}_\mu + \epsilon_{\mn \rho} F_{\mn} V_\rho
-i \epsilon_{\mn \rho} \maruk{ A_\mu \partial_\nu A_\rho - \frac{1}{3} A_\mu [A_\nu, A_\rho] }. 
\end{align}

In the end, we obtain,
\begin{align}
\mathcal{S} &= \mathcal{S}^{B}_0 
+ 2 M^2 V_\mu V_\mu + i M \epsilon_{\mn \rho} V_\mu [V_\nu, V_\rho] \notag \\ 
& ~~~~ - 4 M \lambda \tilde{\lambda} + 4 M \lambda_\mu \tilde{\lambda}_\mu
+ i M \epsilon_{\mn \rho} \maruk{ A_\mu \partial_\nu A_\rho - \frac{1}{3} A_\mu [A_\nu, A_\rho] } 
\end{align}
Potential terms are,
\begin{align}
\mathcal{V}_B &= -\qrt [V_\mu, V_\nu][V_\mu, V_\nu] + 2 M^2 V_\mu V_\mu 
+ i M \epsilon_{\mn \rho} V_\mu [V_\nu, V_\rho].
\end{align}
This is totally different to the previous potentials. 
We redefine the mass parameter as $M \to i M$. Then, the potential is further written as, 
\begin{align}
\mathcal{V}_B&= 
\maruk{\half [V_\mu, V_\nu] - i M \epsilon_{\mn \rho}  V_\rho} \maruk{ -\half [V_\mu, V_\nu] +i M \epsilon_{\mn \sigma} V_\sigma } \notag \\
&= \maruk{\half [V_\mu, V_\nu] - i M \epsilon_{\mn \rho}  V_\rho} \maruk{\half [V_\mu, V_\nu] - i M \epsilon_{\mn \rho}  V_\rho }^\dagger \geq 0.
\end{align}
Hence a classical configuration of the potential satisfies,
\begin{align}
\half [V_\mu, V_\nu] - i M \epsilon_{\mn \rho}  V_\rho &= 0.
\end{align}
This takes the same form as the commutation relation of Lie algebra, so solution can be chosen as, 
\begin{align}
V_\mu &= 2M L_\mu,
\end{align}
where $L_\mu$ ($\mu = 1,2,3$) is $N$-dimensional representation of $SU(2)$, which is a subgroup of gauge group $U(N)$.

In contrast to A-models in the previous sections,  
bosonic vector fields form fuzzy sphere solution in this model and they are proportional to generators of $SU(2)$.

\section{$N=4, D=2$ $U(N)$ SYM} 
We can obtain two dimensional models by dimensional reduction of three dimensional models.
There appear two new scalar supercharges from vector supercharges in three dimensional model,
so that it is similarly possible to consider several combinations of supercharges. 

Before continuing to perform mass deformation, we would evaluate the isometry of two dimensional model.
The isometry of the $N=4, D=2$ model can be obtained 
dimensional reduction from $N=1, D=6$ model. 
In this section we consider the following decomposition of the isometry:
\begin{align}
SO(6) \times SU(2)_R 
\to & SO(2)_E \times  ( SU(2)_N \times SU(2)_R),  
\end{align}
where $SU(2)_R$ is an internal symmetry in $N=1, D=6$ SYM
\footnote{$SO(6)$ can take maximal subgroups as $SO(6) \to SO(4) \times SO(2)$ or $SO(6) \to SO(3) \otimes SO(3)$. 
In the present model we consider the latter decomposition and further take $SO(2)_E$ subgroup of the $SO(3)$}.
Euclidean symmetry is taken as $SO(2)_E$ which is subgroup of $SO(3)_E$.

With the above decomposition,
it is possible to consider two ways of a twist.
\begin{itemize}
\item A-model: Taking diagonal sum of $SO(2)_E$ and $SO(2)_R$ which is subgroup of $SU(2)_R$
\begin{align}
SO(2)_E \times ( SU(2)_N \times SU(2)_R) \to SO(2)_{E'} \times SU(2)_N
\end{align}
In this model, we consider two scalar supercharges transforming as a doublet under $SU(2)_N$.
	
\item B-model: Taking diagonal sum of $SO(2)_E$ with $SO(2)_N$ which is subgroup of $SU(2)_N$
\begin{align}
SO(2)_E \times (SU(2)_N \times SU(2)_R) \to SO(2)_{E'} \times SU(2)_R
\end{align}
In this model, we consider two supercharges transforming as a doublet under $SU(2)_R$.
\end{itemize}
From these consideration we conclude that the internal symmetries in each model are the same one in three dimensional models, 
even though new scalar supercharges appear. 

\subsection{A-model} 
Two dimensional A-twisted action is,
\begin{align}
\mathcal{S}_0 = \int d^2 x \text{Tr}
&\left( \half F_{\mn} F_{\mn} + \mathcal{D}_\mu A_3 \mathcal{D}_\mu A_3 + \mathcal{D}_\mu N \mathcal{D}_\mu N 
-[A_3, N] [A_3, N] \right. \notag \\
& + \mathcal{D}_\mu \phi \mathcal{D}_\mu \overline{\phi} - [A_3, \phi][A_3, \overline{\phi}] 
- [N, \overline{\phi}][N, \phi] + \qrt [\phi, \overline{\phi}]^2 \notag \\
& -i \psi_\mu \mathcal{D}_\mu \psi - \psi_3 [A_3, \psi] -i\chi_\mu \mathcal{D}_\mu \chi - \chi_3[A_3, \chi] \notag \\
& -i \emn \psi_\mu \mathcal{D}_\nu \chi_3 - i \emn \psi_3 \mathcal{D}_\mu \chi_\nu
+ \emn \psi_\mu [A_3, \chi_\nu] \notag \\
& - H_\mu H_\mu - H_3 H_3 
+ \halfi \overline{\phi} \{ \psi_\mu, \psi_\mu \} + \halfi \overline{\phi} \{\psi_3, \psi_3 \} \notag \\
& + \halfi \phi \{ \chi_\mu, \chi_\mu \} + \halfi \phi \{ \chi_3, \chi_3 \} 
+ \halfi \overline{\phi} \{ \chi, \chi \} + \halfi \phi \{ \psi, \psi \} \notag \\
& \left. + N \{\psi, \chi \} + N \{ \psi_\mu, \chi_\mu \} + N \{ \psi_3, \chi_3 \} \right),
\end{align}
where $\mu = 1,2$.

In this model, there are four scalar supercharges $(s, \overline{s}, s_3, \overline{s}_3)$ 
possessing ghost numbers as $(+, -, -, +)$ respectively. 
Then, the action can be expressed as supercharge exact form in four different ways:
\begin{align}
\mathcal{S}_0 
&= \int d^2 x \text{Tr} s \overline{s}
	\maruk{ \chi \psi + \psi_\mu \chi_\mu + \psi_3 \chi_3 + 2i \emn A_3 F_{\mn}} \\
&= \int d^2 x \text{Tr} \overline{s}_3 s_3 
	\maruk{ \chi \psi - \psi_\mu \chi_\mu + \psi_3 \chi_3 - 2i \emn A_3 F_{\mn}} \\
&= \int d^2 x \text{Tr} s s_3 
	\maruk{ \psi_3 \psi + \chi_3 \chi + \emn \psi_\mu \chi_\nu -2i \emn N F_{\mn}} \\
&= \int d^2 x \text{Tr} \overline{s}_3 \overline{s}
	\maruk{ -\psi_3 \psi - \chi_3 \chi + \emn \psi_\mu \chi_\nu - 2i \emn N F_{\mn} }.
\end{align}
In this model all the integrands include BF-term which corresponds to 
Chern-Simons term in three dimensions.
Performing the mass deformation, we can take four different choices of scalar supercharges.
It can be shown that these procedures are
all equivalent by using some discrete symmetries shown in Appendix A.2.
As in the case of $N=8, D=3$ SYM model, there are some discrete symmetries in the above action.
The model, however, possesses a cyclic symmetry, for example supercharges transform as $s \to s_3$, 
$s_3 \to -\overline{s}_3$, $\overline{s}_3 \to \overline{s}$ and $\overline{s} \to s$. From the discussion 
in the beginning of this section, the internal symmetry is $SU(2)$ after the twist. However the internal symmetry might be 
enhanced, because the cyclic symmetry implies that these supercharges belong to the same multiplet in a large group.
As we mention in the previous footnote, maximal subgroup of $SO(6)$ is $SO(2) \otimes SU(2) \otimes SU(2)$.
If this $SO(2)$ symmetry is identified with two dimensional Euclidean one,
the internal symmetry becomes $SO(4)$ after the twist.

Since these procedures of the mass deformation are equivalent, it is sufficient to consider the case of $(s, \overline{s})$ pair.
The simplest way to obtain the result is to perform dimensional reduction of deformed $N=4, D=3$ SYM model.
The results are similar to that of the three dimensional model, and therefore we omit details here.

%

\subsection{B-model}
Two dimensional B-model action is
\begin{align}
\mathcal{S}_0 = \int d^2 x \text{Tr}
& \left( \qrt F_{\mn} F_{\mn} + \half \mathcal{D}_\mu A_3 \mathcal{D}_\mu A_3 
+ \half \mathcal{D}_\mu V_\nu \mathcal{D}_\mu V_\nu + \half \mathcal{D}_\mu V_3 \mathcal{D}_\mu V_3 \right. \notag \\
& - \half [A_3, V_\mu][A_3, V_\mu] - \half [A_3, V_3][A_3,V_3] \notag \\
& -2i \lambda_\mu \mathcal{D}_\mu \tilde{\lambda} -2 \lambda_3 [A_3,\tilde{\lambda}]
-2i \tilde{\lambda}_\mu \mathcal{D}_\mu \lambda -2 \tilde{\lambda}_3 [A_3, \lambda] \notag \\
& -2i \emn \lambda_\mu \mathcal{D}_\nu \tilde{\lambda}_3 -2i \emn \lambda_3 \mathcal{D}_\mu \tilde{\lambda}_\nu
+ 2i \emn \lambda_\mu [A_3, \tilde{\lambda}_\nu] \notag \\
& + 2i V_\mu \{ \lambda_\mu, \tilde{\lambda} \} + 2i V_3 \{ \lambda_3, \tilde{\lambda} \} 
+ 2i V_\mu \{ \tilde{\lambda}_\mu, \lambda \} + 2i V_3 \{ \tilde{\lambda}_3, \lambda \} \notag \\
& + 2i \emn V_\mu \{ \lambda_\nu, \tilde{\lambda}_3 \} -2i \emn V_\mu \{ \lambda_3, \tilde{\lambda}_\nu \}
+ 2i \emn V_3 \{ \lambda_\mu, \tilde{\lambda}_\nu \} \notag \\
& \left. - \qrt [V_\mu, V_\nu]^2 - \half [V_\mu, V_3][V_\mu, V_3] + 2K^2 + 2G \tilde{G} \right)
\end{align}

Then, the exact form of the action is
\begin{align}
\mathcal{S}_0 
&= \int d^2 x \text{Tr} s \overline{s} 
	\maruk{ 2 \tilde{\lambda} \lambda  +2 \tilde{\lambda}_\mu \lambda_\mu  +2  \tilde{\lambda}_3 \lambda_3
	+ \emn \maruk{ V_3 F_{\mn} -i A_3 F_{\mn} + 2 V_\mu \mathcal{D}_\nu A_3} }\\
&= \int d^2 x \text{Tr} \overline{s}_3 s_3 
	\maruk{ 2 \tilde{\lambda} \lambda -2 \tilde{\lambda}_\mu \lambda_\mu + 2 \tilde{\lambda_3} \lambda_3
	+ \emn \maruk{-V_3 F_{\mn} + i A_3 F_{\mn} + 2 V_\mu \mathcal{D}_\nu A_3} } \\
&= \int d^2 x \text{Tr} s s_3
	\maruk{ 2 \lambda \lambda_3 + 2 \tilde{\lambda} \lambda_3 + 2 \emn \tilde{\lambda}_\mu \lambda_\nu + 2i V_3 \mathcal{D}_\mu V_\mu } \\
&= \int d^2 x \text{Tr} \overline{s}_3 \overline{s}
	\maruk{ -2 \lambda \tilde{\lambda}_3 -2 \tilde{\lambda} \lambda_3 + 2 \emn \tilde{\lambda}_\mu \lambda_\nu -2i V_3 \mathcal{D}_\mu V_\mu}
\end{align}
In contrast to the A-model there is no cyclic symmetry. Not four all different procedures of the mass deformation
are not equivalent There are however some discrete symmetries in this model, so that it is sufficient to 
consider $(s, \overline{s})$ and $(s, s_3)$ cases in the deformation. 
A characteristic of this model is that the integrand of $(s, \overline{s})$ includes BF-term and 
that of $(s, s_3)$ does not include it. We will try to perform the deformation of each model.

\subsubsection{$s, \overline{s}$ action}
We define
\begin{align}
Q &= s + \Delta s, & \overline{Q} &= \overline{s} + \Delta \overline{s},
\end{align}
and assume the following algebra,
\begin{align}
Q^2 &= M J_{++}, & \overline{Q}^2 &= -M J_{--}, & \{ Q, \overline{Q} \} &= - M J_0.
\end{align}
We obtain the following transformation law of $\Delta s$ and $\Delta \overline{s}$,
\begin{align}
\Delta s (G) &= 0, & \Delta \overline{s} (G) &= 2 M \lambda, \\
\Delta s (\tilde{G}) &= -2 M \tilde{\lambda}, & \Delta \overline{s} (\tilde{G}) &= 0, \\
\Delta s (K) &= M \lambda, & \Delta \overline{s} (K) &= M \tilde{\lambda}, \\
\Delta s (\lambda_3) &= 0, & \Delta \overline{s} (\lambda_3) &= -M V_3, \\
\Delta s (\tilde{\lambda}_3) &= M V_3, & \Delta \overline{s} (\tilde{\lambda}_3) &= 0, \\
\Delta s (\lambda_\mu) &= 0, & \Delta \overline{s} (\lambda_\mu) &= -M V_\mu, \\
\Delta s (\tilde{\lambda}_\mu) &= M V_\mu, & \Delta \overline{s} (\tilde{\lambda}_\mu) &= 0.
\end{align}

The action is now deformed as, 
\begin{align}
\mathcal{S} &= (Q \overline{Q} - M) \mathcal{F}_0,
\end{align}
where
\begin{align}
\mathcal{F}_0 &=2 \tilde{\lambda} \lambda  +2 \tilde{\lambda}_\mu \lambda_\mu  +2  \tilde{\lambda}_3 \lambda_3
	+ \emn \maruk{ V_3 F_{\mn} -i A_3 F_{\mn} + 2 V_\mu \mathcal{D}_\nu A_3}.
\end{align}
Then, the deformed action is,
\begin{align}
\mathcal{S} &= \mathcal{S}_0 + 2M \maruk{ \tilde{\lambda} \lambda - 2\tilde{\lambda}_\mu \lambda_\mu - 2\tilde{\lambda}_3 \lambda_3 } \notag \\
& ~~  + 3i M \emn V_\mu [V_\nu, A_3] +i M \emn A_3 F_{\mn} 
+ 2 M^2 (V_\mu V_\mu + V_3 V_3).
\end{align}
The potential terms which involve $V_\mu$ and $V_3$ are,
\begin{align}
V &= - \qrt [V_\mu, V_\nu]^2 - \half [V_\mu, V_3][V_\mu, V_3] 
	+ 3i M \emn V_\mu [V_\nu, V_3] + 2 M^2 (V_\mu V_\mu + V_3 V_3).
\end{align}
These results are exactly the same as those which are obtained 
through the dimensional reduction of three dimensional B-model.
A vacuum solution can be obtained in the similar way, and therefore 
we have to identify $V_\mu$ and $V_3$ with $SU(2)$ generators to form the fuzzy sphere solution.

\subsubsection{$s, s_3$ action (same for $\overline{s}_3, \overline{s}$)}
In this combination, we can not find any descriptions of supertransformation of deformed charges.
The cause of this problem has not been cleared yet.
This problem might be related to the internal symmetry in these models.
From the discussion of the discrete symmetry A-model has larger internal symmetry than that of B-model.
However, it is not clear that this fact affects whether the mass deformation is possible.

\section{Summary and discussion} 
In this paper, we have evaluated whether the mass deformation method, introduced in \cite{Sug} is applicable to any SYM models 
and also focused on the fuzzy sphere solution to resolve the flat direction which typically appear in higher $N$-extended SYM models.

We have found that the key ingredients of mass deformation are 
that the theory should have at least two topological supercharges and their ghost numbers
should be opposite. Therefore we have focused on the models which satisfy both characteristics: 
maximal SUSY or half maximal SUSY models in four dimensions or less. 

The mass deformation procedure is as follow:
\begin{enumerate}
\item Find the combination of supercharges whose ghost numbers are opposite, 
	and write the action in exact form in their supercharges.
\item Deform their supercharges, and define deformed algebra.
\item The action is  exact with deformed supercharges, and is invariant under its two supersymmetries.
\item Through this process, the action receives mass terms, and the potential terms are changed.
\item Classical configuration constrains scalar fields to satisfy the algebra of $SU(2)$. 
	Thus the fuzzy sphere solution appear.
\end{enumerate}

We have studied the following models: 
$N=4, D=4$ SYM, $N=8, D=3$ SYM, two different twisted models of $N=4, D=3$ SYM  and of  $N =(4,4), D=2$ SYM .
We investigate the classical solution of the potential in each case. 
Some of them are connected by the dimensional reduction, so that the result is sometimes straightforward. 
But dimensional reduction produces new scalar supercharges derived from vector supercharge in 
higher dimensions, and it contributes to new combinations of supercharges.
Even though there appear new supercharges and new combinations of them, 
they are connected each other through the discrete symmetries. 
It is therefore sometimes enough to evaluate only one model.  

Since A-twisted models always possess at least one $SU(2)$ $R$-symmetry, it results in the fact that 
the theories possess two nilpotent supercharges which transform as doublet under $SU(2)$ $R$-symmetry. 
Indeed, the models which satisfy above conditions are always possible to perform mass deformation and no exception arises 
with respect to the model we treat in this paper.  

The difference appears in B-twisted models. 
The mass deformation is applicable to $N=4, D=3$ model, and the problem does not arise.
Through the dimensional reduction we obtain $N=(4,4), D=2$ B-model.
But the deformations using some new combinations of supercharges are not allowed, 
because there is no reasonable definition of supertransformation which is consistent with deformed algebra.
This might be caused by insufficiency of $R$-symmetry, 
since the cyclic discrete symmetry is not found in this model. 

There are some points which are not fully clarified, and they are left for future work:
\begin{itemize}
\item The difference of the results in A- and B-models, from the point of internal symmetry.
\item Treatment for complex terms which should be excluded for lattice simulation.
\end{itemize}

\noindent
{\Large \textbf{Acknowledgment}}\\

\noindent
We would like to thank N. Kawamoto, F. Sugino and I. Kanamori for fruitful discussion and comments.
A. M. would like to thank K. Ohta for useful comments.

\appendix

\section{Discrete symmetry}
\subsection{Discrete symmetries in $N=8, D=3$ model} 

\mbox{}

{\bf Symmetry obtained through 4D (ghost number flipping): $(s \leftrightarrow \overline{s}), (\overline{s}_4 \leftrightarrow s_4)$}
\begin{align*}
A_i & \to A_i, & A_4 & \to A_4, & v &\to v, \\
\phi &\to - \overline{\phi}, & \overline{\phi} &\to - \phi,  & v_{i} &\to -v_{i} \\
\chi &\leftrightarrow \lambda, & \chi_{i} &\leftrightarrow \lambda_{i}, 
& C_i &\leftrightarrow \psi_i, & C_4 & \leftrightarrow \psi_4 .
\end{align*}

{\bf ghost number flipping: $( s \to s_4, s_4 \to - s), (\overline{s}_4 \to \overline{s}, \overline{s} \to -\overline{s}_4)$}
\begin{align*}
A_i & \to A_i, & v_{i} & \to v_{i}, &  v & \to v, \\
\phi & \to - \overline{\phi}, & \overline{\phi} & \to - \phi, & A_4 & \to -A_4,  \\
\lambda & \to -\psi_4, & \psi_4 & \to \lambda, & \chi & \to C_4, & C_4 & \to - \chi,  \\
\lambda_i & \to \psi_i, & \psi_i & \to - \lambda_{i}, & \chi_i & \to -C_i, & C_i & \to \chi_{i}. 
\end{align*}

{\bf ghost number conserving: $(s \to \overline{s}_4, \overline{s}_4 \to -s), (\overline{s} \to s_4, s_4 \to - \overline{s})$}
\begin{align*}
A_i & \to A_i,   & v_{i} & \to v_{i}, & \phi & \to \phi, & \overline{\phi} & \to \overline{\phi}, \\
A_4 & \to - A_4, &  v & \to -v, \\
\lambda & \to C_4, & C_4 & \to - \lambda, & \chi & \to \psi_4, & \psi_4 & \to - \chi,  \\
\lambda_i & \to - C_i,  & C_i & \to \lambda_{i}, & \chi_i & \to -\psi_i,  & \psi_i & \to \chi_{i}.
\end{align*}

\subsection{Discrete symmetries in $N=4, D=2$ A-model SYM} 
\mbox{}

{\bf ghost number flipping: ($s \to \overline{s}, \overline{s} \to -s$), ($s_3 \to \overline{s}_3, \overline{s}_3 \to -s_3$)}
\begin{align*}
A_\mu &\to A_\mu, & A_3 & \to A_3, & H_\mu &\to H_\mu, & H_3 &\to H_3, \\
\phi &\leftrightarrow \overline{\phi}, & N &\to -N \\
\psi_\mu &\to \chi_\mu, & \psi &\to \chi, & \psi_3 & \to \chi_3, \\
\chi_\mu &\to -\psi_\mu, & \chi &\to - \psi, & \chi_3 & \to -\psi_3.
\end{align*}

{\bf ghost number flipping: ($s \to s_3, s_3 \to -s$), ($\overline{s}_3 \to \overline{s}, \overline{s} \to - \overline{s}_3$)}
\begin{align*}
A_\mu &\to A_\mu, & N &\to N, & H_\mu &\to H_\mu, & H_3 &\to H_3 \\
\phi &\leftrightarrow \overline{\phi}, & A_3 &\to - A_3, \\
\psi_\mu & \to \emn \chi_\nu, & \psi& \to \psi_3, & \chi_3 & \to \chi, \\
\chi_\mu & \to \emn \psi_\nu, & \psi_3 &\to - \psi, & \chi & \to - \chi_3.
\end{align*}

{\bf ghost number conserving: ($s \to \overline{s}_3, \overline{s}_3 \to -s$), ($s_3 \to \overline{s}, \overline{s} \to - s_3$)}
\begin{align*}
A_\mu &\to A_\mu, & A_3 &\to A_3, & N &\to N, & H_3 &\to H_3, \\
\phi &\to \phi, & \overline{\phi} & \to \overline{\phi}, & H_\mu & \to - H_\mu, \\
\psi_\mu &\to - \emn \psi_\nu, & \psi &\to \chi_3, & \psi_3 &\to \chi, \\
\chi_\mu &\to - \emn \chi_\nu, & \chi_3 &\to - \psi, & \chi &\to - \psi_3.
\end{align*}

{\bf ghost number flipping forward cyclic: ($\overline{s}_3 \to \overline{s} \to s \to s_3 \to -\overline{s}_3$)}
\begin{align*}
A_\mu &\to A_\mu, & H_3 &\to H_3, \\
\phi & \leftrightarrow \overline{\phi}, & A_3 & \leftrightarrow N,  & H_\mu &\to - \emn H_\nu, \\
\psi &\to \psi_3, & \psi_3 &\to - \chi_3, & \chi_3 &\to \chi, & \chi &\to \psi, \\
\psi_\mu & \to \emn \chi_\nu, & \chi_\mu &\to \psi_\mu.
\end{align*}

{\bf ghost number flipping backward cyclic: ($s_3 \to s \to \overline{s} \to \overline{s}_3 \to - s_3$)}
\begin{align*}
A_\mu &\to A_\mu, & H_3 &\to H_3, \\
\phi & \leftrightarrow \overline{\phi}, & A_3 & \leftrightarrow N,  & H_\mu &\to \emn H_\nu, \\
\psi & \to \chi, & \chi & \to \chi_3, & \chi_3 & \to - \psi_3, & \psi_3 & \to \psi, \\
\psi_\mu & \to \chi_\nu, & \chi_\mu &\to - \emn \psi_\mu.
\end{align*}

\subsection{Discrete symmetries in $N=4, D=2$ B-model SYM}
\mbox{}

{\bf ghost number flipping: ($s \to \overline{s}, \overline{s} \to -s$), ($s_3 \to \overline{s}_3, \overline{s}_3 \to -s_3$)}
\begin{align*}
V_\mu &\to V_\mu, & V_3 &\to V_3, & A_\mu &\to A_\mu, & A_3 & \to A_3, \\
G & \to - \tilde{G}, & \tilde{G} & \to - G, & K & \to -K, \\
\lambda &\to - \tilde{\lambda}, & \lambda_\mu & \to \tilde{\lambda}_\mu, & \lambda_3 &\to \tilde{\lambda}_3, \\
\tilde{\lambda} &\to \lambda, & \tilde{\lambda}_\mu & \to - \lambda_\mu, & \tilde{\lambda}_3 & \to - \lambda_3.
\end{align*}

{\bf ghost number flipping: ($s \to s_3, s_3 \to -s$), ($\overline{s}_3 \to \overline{s}, \overline{s} \to - \overline{s}_3$)}
\begin{align*}
V_\mu &\to - V_\mu, & V_3 &\to -V_3, & A_\mu &\to A_\mu, & A_3 &\to -A_3, \\
G &\to  \tilde{G}, & \tilde{G} &\to G, & K &\to -K, \\
\lambda &\to - \tilde{\lambda}_3, & \tilde{\lambda} &\to \lambda_3, & \lambda_\mu &\to \emn \tilde{\lambda}_\nu, \\
\tilde{\lambda}_3, &\to \lambda, & \lambda_3 &\to -\tilde{\lambda}, & \tilde{\lambda}_\mu &\to \emn \lambda_\nu.
\end{align*}

{\bf ghost number conserving: ($s \to \overline{s}_3, \overline{s}_3 \to -s$), ($s_3 \to \overline{s}, \overline{s} \to - s_3$)}
\begin{align*}
V_\mu &\to - V_\mu, & V_3 &\to V_3, & A_\mu &\to A_\mu, & A_3 &\to A_3, \\
G &\to G, & \tilde{G} &\to \tilde{G}, & K &\to K, \\
\lambda &\to - \lambda_3, & \tilde{\lambda} &\to \tilde{\lambda}_3, & \lambda_\mu &= - \emn \lambda_\nu, \\
\lambda_3 &\to \lambda, & \tilde{\lambda}_3 &= - \tilde{\lambda}, & \tilde{\lambda}_\mu &= - \emn \tilde{\lambda}_\nu.
\end{align*}

\section{Tables of supertransformation}
\subsection{Supertransformation in $N=4, D=4$} 

\begin{tabular}{|c|c||l|l|}
\hline
 & gh\# & $s$ & $\overline{s}$ \\ \hline
 $v$ & $0$ & $\half \lambda$ & $\half \chi$ \\
 $v_B^+$ & $0$ & $-\half \lambda_B^+$ & $\half \chi_B^+$ \\
 $\lambda$ & $1$ & $-i [\phi, v]$ & -$\halfi [\phi, \overline{\phi}] $ \\
 $\lambda_B^+$ & $1$ & $ i [\phi, v_B^+] $ & $\frac{i}{32} \Gamma^+_{BCD} [v^+_C, v^+_D] + i [v_B^+, v] -2 F_B^+ + h_B^+$ \\ 
 $\psi_\mu$ & $-1$ & $i \mathcal{D}_\mu v + i \mathcal{D}_\rho v_{\mu \rho}^+ + h_\nu$ & $i \mathcal{D}_\mu \overline{\phi}$ \\
 $\phi$ & $2$ & $0$ & $\lambda$ \\
 $\overline{\phi}$ & $-2$ & $-\chi$ & $0$ \\
 $\chi$ & $-1$ & $\halfi [\phi, \overline{\phi}]$ & $i [\overline{\phi}, v]$ \\
 $\chi_B^+$ & $-1$ & $\frac{i}{32} \Gamma^+_{BCD} [v^+_C, v^+_D] -i [v_B^+, v] - 2 F_B^+ + h^+_B$ & $i[\overline{\phi}, v_B^+] $ \\
 $C_\mu$ & $1$ & $-i \mathcal{D}_\mu \phi$ & $i \mathcal{D}_\mu v - i \mathcal{D}_\rho v^+_{\mu \rho} - h_\mu$ \\
 $A_\mu$ & $0$ & $-\halfi C_\mu$ & $-\halfi \psi_\mu$ \\ \hline
$h_\mu$ & $0$ & 
\begin{tabular}{l} $-\halfi \mathcal{D}_\mu \lambda + \halfi \mathcal{D}_\rho \lambda^+_{\mu \rho} $ \\
$+ \halfi [C_\rho, v^+_{\mu \rho}] - \halfi [v, C_\mu] - \halfi [\phi, \psi_\mu] $\end{tabular} &
\begin{tabular}{l} $\halfi \mathcal{D}_\mu \chi -\halfi \mathcal{D}_\rho \chi^+_{\mu \rho}$ \\ 
$ + \halfi [\psi_\rho, v^+_{\mu \rho}] + \halfi [v, \psi_\mu] - \halfi [\overline{\phi}, C_\mu] $ \end{tabular} \\ 
 & & & \\
$h^+_B$ & $0$ &  
\begin{tabular}{l} $ - \halfi [ \phi ,  \chi^+_B ] + \halfi [v, \lambda^+_B] + \halfi [v^+_B, \lambda]$ \\
    $-\halfi \delta_{B, \mn}^+ \mathcal{D}_\mu C_\nu + \frac{i}{32} \Gamma^+_{BCD} [\lambda^+_C, v^+_D]   $\end{tabular} &
\begin{tabular}{l} $\halfi [\overline{\phi}, \lambda^+_B]  + \halfi [v, \chi^+_B]  - \halfi [v^+_B, \chi] $\\
    $-\halfi \delta_{B, \mn}^+ \mathcal{D}_\mu \psi_\nu -\frac{i}{32} \Gamma^+_{BCD} [\chi^+_C, v^+_D ]$ \end{tabular} \\
\hline 
\end{tabular}\\

\subsection{Supertransformation in $N=8, D=3$} 
{\bf ($s, \overline{s}$) supertransformation}\\

\begin{tabular}{|c|c||l|l|}
\hline
 & gh\# & $s$ ~~(gh\# $=1$) & $\overline{s}$ ~~(gh\# $=-1$) \\ \hline
 $v$ & $0$ & $\half \lambda$ &  $\half \chi$ \\
 $v_{i}$ & $0$ & $-\half \lambda_{i}$ &  $\half \chi_{i}$  \\
 $\lambda$ & $1$ & $-i [\phi, v]$ & $-\halfi [\phi, \overline{\phi}] $ \\
 $\lambda_{i}$ & $1$ & $ i [\phi, v_{i}] $ 
	& \begin{tabular}{l} $-\frac{i}{2} \eijk [v_{j}, v_{k}] + i [v_{i}, v]  $ \\ 
		\hspace*{10mm} $- \mathcal{D}_i A_4 + \half \eijk F_{jk} + h_{i4}^+$ \end{tabular} \\ 
 $\psi_i$ & $-1$ & $i \mathcal{D}_i v - i \eijk \mathcal{D}_j v_{k} + [A_4, v_{i}] + h_i$ &  $i \mathcal{D}_i \overline{\phi}$ \\
 $\psi_4$ & $-1$ & $[A_4, v] - i \mathcal{D}_j v_{j} + h_4$ &  $[A_4, \overline{\phi}]$ \\
 $\phi$ & $2$ & $0$ & $\lambda$ \\
 $\overline{\phi}$ & $-2$ & $-\chi$ & $0$ \\
 $\chi$ & $-1$ & $\halfi [\phi, \overline{\phi}]$ & $i [\overline{\phi}, v]$ \\
 $\chi_{i}$ & $-1$ 
	& \begin{tabular}{l} $-\frac{i}{2} \eijk [v_{j}, v_{k}] -i [v_{i}, v] $ \\ 
		\hspace*{10mm} $- \mathcal{D}_i A_4 + \half \eijk F_{jk} + h^+_{i4}$ \end{tabular} & $i[\overline{\phi}, v_{i}] $ \\
 $C_i$ & $1$ & $-i \mathcal{D}_i \phi$ &  $i \mathcal{D}_i v + i \eijk \mathcal{D}_j v_k - [A_4, v_i] - h_i$  \\
 $C_4$ & $1$ & $-[A_4, \phi]$ &  $[A_4, v] + i \mathcal{D}_j v_{j} - h_4$ \\
 $A_i$ & $0$ & $-\halfi C_i$ &  $-\halfi \psi_i$ \\
 $A_4$ & $0$ & $-\halfi C_4$ &  $-\halfi \psi_4$ \\  \hline
\end{tabular}\\

\noindent
{\bf Transformation of auxiliary fields by $s$ and $\overline{s}$} \\ 

\begin{tabular}{|c||l|l|}
\hline
& $s$ & $\overline{s}$ \\ \hline
& & \\
$h^+_{i4}$ &
\begin{tabular}{l} $-\halfi [\phi, \chi_{i}] - \halfi [\lambda_{i}, v] + \halfi [ v_{i}, \lambda] - \half [C_i, A_4]$ \\
$- \halfi \eijk [\lambda_{j}, v_{k}] - \halfi \mathcal{D}_i C_4 + \halfi \eijk \mathcal{D}_j C_k$ \end{tabular}
& \begin{tabular}{l} $\halfi[\overline{\phi}, \lambda_{i}] - \halfi [ \chi_{i}, v] - \halfi [v_{i}, \chi] - \half [\psi_i, A_4]$ \\
$+ \halfi \eijk [\chi_{i}, v_{k}] - \halfi \mathcal{D}_i \psi_4 + \halfi \eijk \mathcal{D}_j \psi_k$ \end{tabular} \\
& & \\
$h_i$ & 
\begin{tabular}{l} $ -\halfi [\phi, \psi_i] + \halfi [C_i, v]  + \halfi [C_4, v_{i}] + \half [A_4, \lambda_{i}]$ \\
$- \halfi \eijk [C_j, v_{k}] - \halfi \eijk \mathcal{D}_j \lambda_{k} - \halfi \mathcal{D}_i \lambda $ \end{tabular} 
&\begin{tabular}{l} $-\halfi [\overline{\phi}, C_i] - \halfi [\psi_i, v] + \halfi [ \psi_4, v_{i}] - \half [A_4, \chi_{i}]$ \\
$- \halfi \eijk [ \psi_j, v_{k}] + \halfi \eijk \mathcal{D}_j \chi_{k} + \halfi \mathcal{D}_i \chi $ \end{tabular}  \\
& & \\
$h_4$ &
\begin{tabular}{l} 
$-\halfi [\phi, \psi_4] + \halfi [C_4, v] -\half [A_4, \lambda] - \halfi [C_j, v_{j}]$ \\ $ -\halfi \mathcal{D}_j \lambda_{j}$ 
\end{tabular} 
& \begin{tabular}{l} 
$-\halfi [\overline{\phi}, C_4] - \halfi [\psi_4, v] + \half [A_4, \chi] - \halfi [\psi_j, v_{j}] $ \\ $+ \halfi \mathcal{D}_j \chi_{j}$
\end{tabular}\\ \hline
\end{tabular}\\

\newpage
\noindent
{\bf ($\overline{s}_4, s_4$) supertransformation}\\ 

\begin{tabular}{|c|c||l|l|l|}
\hline
 & gh\# & $\overline{s}_4$ ~~(gh\# $=1$) & $s_4$ ~~(gh\# $=-1$) \\ \hline
 $v$ & $0$ & $-\half C_4$ & $- \half \psi_4$ \\
 $v_{i}$ & $0$ &  $\half C_i$ & $-\half \psi_i$ \\
 $\lambda$ & $1$ &  $-[A_4, \phi] $ & $-[A_4, v] - i \mathcal{D}_j v_{j} + g$ \\
 $\lambda_{i}$ & $1$ & $-i \mathcal{D}_i \phi$ & $-i \mathcal{D}_i v + [A_4, v_{i}] + i \eijk \mathcal{D}_j v_{k} + g^+_{i4}$ \\ 
 $\psi_i$ & $-1$ 
	& \begin{tabular}{l} $\frac{i}{2} \eijk [ v_{j}, v_{k} ] - i [ v_{i}, v] $ \\ 
		\hspace*{10mm} $ - \mathcal{D}_i A_4 - \half \eijk F_{jk} + g_i$ \end{tabular} & $- i [ \overline{\phi}, v_{i}] $  \\
 $\psi_4$ & $-1$ & $\halfi [\phi, \overline{\phi}] $ & $-i [\overline{\phi}, v]  $ \\
 $\phi$ & $2$ & $0$ & $C_4$ \\
 $\overline{\phi}$ & $-2$ & $- \psi_4$ & $0$ \\
 $\chi$ & $-1$ &  $-[A_4, v] + i \mathcal{D}_j v_{j} - g$ & $[A_4, \overline{\phi}] $ \\
 $\chi_{i}$ & $-1$ & $-i \mathcal{D}_i v - [A_4, v_{i}] - i \eijk \mathcal{D}_j v_{k} - g^+_{i4}$ & $i \mathcal{D}_i \overline{\phi}$ \\
 $C_i$ & $1$ & $ -i [\phi, v_{i} ] $ 
	& \begin{tabular}{l} $\frac{i}{2} \eijk [v_{j}, v_{k}] + i [v_{i}, v] $ \\ 
		\hspace*{10mm} $ - \mathcal{D}_i A_4 - \half \eijk F_{jk} + g_i$ \end{tabular} \\
 $C_4$ & $1$ & $i [\phi, v] $ & $ - \halfi  [\phi, \overline{\phi}] $ \\
 $A_i$ & $0$ &  $-\halfi \lambda_{i}$ & $-\halfi \chi_{i}$ \\
 $A_4$ & $0$ &  $-\halfi \lambda$ & $-\halfi \chi $ \\ \hline
 \end{tabular}\\

\noindent
{\bf Transformation of auxiliary fields by $\overline{s}_4$ and $s_4$}\\ 

\begin{tabular}{|c||l|l|}
\hline
& $\overline{s}_4$ &  $s_4$  \\ \hline
& & \\
$g^+_{i4}$ & 
\begin{tabular}{l} $\halfi [\phi, \chi_{i}] + \halfi [\lambda_{i}, v] + \halfi [\lambda, v_{i}] - \half [A_4, C_i]$ \\
$+ \halfi \mathcal{D}_i C_4 + \halfi \eijk [ \lambda_{j}, v_{k}] - \halfi \eijk \mathcal{D}_j C_k$ \end{tabular}
& \begin{tabular}{l} $\halfi [\overline{\phi}, \lambda_{i}] - \halfi [\chi_{i}, v] + \halfi [ \chi, v_{i}] + \half [A_4, \psi_i]$ \\
$- \halfi \mathcal{D}_i \psi_4 + \halfi \eijk [\chi_{j}, v_{k}] + \halfi \eijk \mathcal{D}_j \psi_k$ \end{tabular} \\
& & \\
$g_i$
& 
\begin{tabular}{l} $-\halfi [\phi, \psi_i] + \halfi [C_i, v] - \halfi [v_{i}, C_4] - \half [\lambda_{i}, A_4]$ \\
$- \halfi \eijk [C_j, v_{k}] - \halfi \mathcal{D}_i \lambda - \halfi \eijk \mathcal{D}_j \lambda_{k}$ \end{tabular}
& \begin{tabular}{l} $\halfi [\overline{\phi}, C_i] + \halfi [\psi_i, v] + \halfi [ v_{i}, \psi_4] - \half [\chi_{i}, A_4]$ \\
$+\halfi \eijk [\psi_j, v_{k}] - \halfi \mathcal{D}_i \chi - \halfi \eijk \mathcal{D}_j \chi_{k}$ \end{tabular} \\
& & \\
$g$ 
&
\begin{tabular}{l} $\halfi [\phi, \chi] + \halfi [\lambda, v] + \half [A_4, C_4] - \halfi [ \lambda_{j}, v_{j}]$ \\
$+ \halfi \mathcal{D}_j C_j$ \end{tabular}
& \begin{tabular}{l} $\halfi [ \overline{\phi}, \lambda] - \halfi [\chi, v] - \half [A_4, \psi_4] - \halfi [\chi_{j}, v_{j}]$ \\
$- \halfi \mathcal{D}_j \psi_j$ \end{tabular}  \\ \hline
\end{tabular}\\

\newpage
\noindent
{\bf Supertransformation of $(s, s_4)$ } \\ 


\begin{tabular}{|c|c||l|l|}
\hline
 & gh\# & $s$ ~~(gh\# $=1$) & $s_4$ ~~(gh\# $=-1$)  \\ \hline
 $v$ & $0$ & $\half \lambda$ & $- \half \psi_4$\\
 $v_{i}$ & $0$ & $-\half \lambda_{i}$ & $-\half \psi_i$ \\
 $\lambda$ & $1$ & $-i [\phi, v]$ & $-[A_4, v] - i \mathcal{D}_j v_{j} + L$ \\
 $\lambda_{i}$ & $1$ & $ i [\phi, v_{i}] $ & $-i \mathcal{D}_i v + [A_4, v_{i}] + i \eijk \mathcal{D}_j v_{k} -N_i$ \\ 
 $\psi_i$ & $-1$ & $i \mathcal{D}_i v - i \eijk \mathcal{D}_j v_{k} + [A_4, v_{i}] + N_i$ & $- i [ \overline{\phi}, v_{i}] $ \\
 $\psi_4$ & $-1$ & $[A_4, v] - i \mathcal{D}_j v_{j} + L$ & $-i [\overline{\phi}, v]  $ \\
 $\phi$ & $2$ & $0$ & $C_4$ \\
 $\overline{\phi}$ & $-2$ & $-\chi$ & $0$  \\
 $\chi$ & $-1$ & $\halfi [\phi, \overline{\phi}]$ & $[A_4, \overline{\phi}] $ \\
 $\chi_{i}$ & $-1$ 
	& \begin{tabular}{l} $-\frac{i}{2} \eijk [v_{j}, v_{k}] -i [v_{i}, v] $ \\ 
	\hspace*{10mm} $- \mathcal{D}_i A_4 + \half \eijk F_{jk} + M_i$ \end{tabular} & $i \mathcal{D}_i \overline{\phi}$ \\
 $C_i$ & $1$ & $-i \mathcal{D}_i \phi$ 
	& \begin{tabular}{l} $\frac{i}{2} \eijk [v_{j}, v_{k}] + i [v_{i}, v] $ \\ 
	\hspace*{10mm} $- \mathcal{D}_i A_4 - \half \eijk F_{jk} -M_i$ \end{tabular} \\
 $C_4$ & $1$ & $-[A_4, \phi]$ & $ - \halfi  [\phi, \overline{\phi}] $ \\
 $A_i$ & $0$ & $-\halfi C_i$ & $-\halfi \chi_{i}$ \\
 $A_4$ & $0$ & $-\halfi C_4$ & $-\halfi \chi $ \\  \hline
\end{tabular}\\

\noindent
{\bf Transformation of auxiliary fields in $(s, s_4)$ case}\\ 

\begin{tabular}{|c||l|l|}
\hline
& $s$ &  $s_4$ \\ \hline
& & \\
$M_i$ & 
\begin{tabular}{l} $-\halfi [\phi, \chi_{i}] - \halfi [\lambda_{i}, v] + \halfi [ v_{i}, \lambda] - \half [C_i, A_4]$ \\
$- \halfi \eijk [\lambda_{j}, v_{k}] - \halfi \mathcal{D}_i C_4 + \halfi \eijk \mathcal{D}_j C_k$ \end{tabular}
& \begin{tabular}{l} $-\halfi [\overline{\phi}, C_i] - \halfi [\psi_i, v] - \halfi [ v_{i}, \psi_4] + \half [\chi_{i}, A_4]$ \\
$- \halfi \eijk [\psi_j, v_{k}] + \halfi \mathcal{D}_i \chi + \halfi \eijk \mathcal{D}_j \chi_{k}$ \end{tabular} 
 \\
& & \\
$N_i$ &
\begin{tabular}{l} $ -\halfi [\phi, \psi_i] + \halfi [C_i, v]  + \halfi [C_4, v_{i}] + \half [A_4, \lambda_{i}]$ \\
$- \halfi \eijk [C_j, v_{k}] - \halfi \eijk \mathcal{D}_j \lambda_{k} - \halfi \mathcal{D}_i \lambda $ \end{tabular}
& \begin{tabular}{l} $-\halfi [\overline{\phi}, \lambda_{i}] + \halfi [\chi_{i}, v] - \halfi [ \chi, v_{i}] - \half [A_4, \psi_i]$ \\
$+ \halfi \mathcal{D}_i \psi_4 - \halfi \eijk [\chi_{j}, v_{k}] - \halfi \eijk \mathcal{D}_j \psi_k$ \end{tabular}
 \\
& & \\
$L$ &
\begin{tabular}{l} 
$-\halfi [\phi, \psi_4] + \halfi [C_4, v] -\half [A_4, \lambda] - \halfi [C_j, v_{j}]$ \\ $ -\halfi \mathcal{D}_j \lambda_{j}$ \end{tabular} 
&\begin{tabular}{l} $\halfi [ \overline{\phi}, \lambda] - \halfi [\chi, v] - \half [A_4, \psi_4] - \halfi [\chi_{j}, v_{j}]$ \\
$- \halfi \mathcal{D}_j \psi_j$ \end{tabular}  \\ \hline
\end{tabular}\\

\newpage
\noindent
{\bf Supertransformation of $ (\overline{s}, \overline{s}_4)$}\\ 

\begin{tabular}{|c|c||l|l|}
\hline
 & gh\# & $\overline{s}_4$ ~~(gh\# $=1$) & $\overline{s}$ ~~(gh\# $=-1$) \\ \hline
 $v$ & $0$ & $-\half C_4$ &  $\half \chi$ \\
 $v_{i}$ & $0$ &  $\half C_i$ &  $\half \chi_{i}$ \\
 $\lambda$ & $1$ &  $-[A_4, \phi] $ & -$\halfi [\phi, \overline{\phi}] $  \\
 $\lambda_{i}$ & $1$ & $-i \mathcal{D}_i \phi$ 
	& \begin{tabular}{l} $- \halfi \eijk [v_{j}, v_{k}] + i [v_{i}, v] $\\ 
	\hspace*{10mm} $- \mathcal{D}_i A_4 + \half \eijk F_{jk} + M_i$ \end{tabular} \\ 
 $\psi_i$ & $-1$ 
	& \begin{tabular}{l} $\frac{i}{2} \eijk [ v_{j}, v_{k} ] - i [ v_{i}, v] $\\
	\hspace*{10mm} $ - \mathcal{D}_i A_4 - \half \eijk F_{jk} - M_i$ \end{tabular} & $i \mathcal{D}_i \overline{\phi}$ \\
 $\psi_4$ & $-1$ & $\halfi [\phi, \overline{\phi}] $ & $[A_4, \overline{\phi}]$ \\
 $\phi$ & $2$ & $0$ & $\lambda$ \\
 $\overline{\phi}$ & $-2$ & $- \psi_4$ & $0$ \\
 $\chi$ & $-1$ &  $-[A_4, v] + i \mathcal{D}_k v_{k} - L$ & $i [\overline{\phi}, v]$ \\
 $\chi_{i}$ & $-1$ & $-i \mathcal{D}_i v - [A_4, v_{i}] - i \eijk \mathcal{D}_j v_{k} + N_i$ & $i[\overline{\phi}, v_{i}] $ \\
 $C_i$ & $1$ & $ -i [\phi, v_{i} ] $ & $i \mathcal{D}_i v + i \eijk \mathcal{D}_j v_{k} - [A_4, v_{i}] - N_i$ \\
 $C_4$ & $1$ & $i [\phi, v] $ & $[A_4, v] + i \mathcal{D}_j v_{j} - L$ \\
 $A_i$ & $0$ &  $-\halfi \lambda_{i}$ & $-\halfi \psi_i$ \\
 $A_4$ & $0$ &  $-\halfi \lambda$ & $-\halfi \psi_4$ \\ \hline
\end{tabular}\\

\noindent
{\bf Transformation of auxiliary fields in $(\overline{s}_4, \overline{s})$ }\\ 

\begin{tabular}{|c||l|l|}
\hline
& $\overline{s}_4$ &  $\overline{s}$ \\ \hline
& & \\
$M_i$ 
& \begin{tabular}{l} $ \halfi [\phi, \psi_i] - \halfi [C_i, v] + \halfi [v_{i}, C_4] + \half [\lambda_{i}, A_4]$ \\
$+ \halfi \eijk [C_j, v_{k}] + \halfi \mathcal{D}_i \lambda + \halfi \eijk \mathcal{D}_j \lambda_{k}$ \end{tabular} 
& \begin{tabular}{l} $\halfi[\overline{\phi}, \lambda_{i}] - \halfi [ \chi_{i}, v] - \halfi [v_{i}, \chi] - \half [\psi_i, A_4]$ \\
$+ \halfi \eijk [\chi_{i}, v_{k}] - \halfi \mathcal{D}_i \psi_4 + \halfi \eijk \mathcal{D}_j \psi_k$ \end{tabular} \\
& & \\
$N_i$  
&  \begin{tabular}{l} $- \halfi [\phi, \chi_{i}] - \halfi [\lambda_{i}, v] - \halfi [\lambda, v_{i}] + \half [A_4, C_i]$ \\
$- \halfi \mathcal{D}_i C_4 - \halfi \eijk [ \lambda_{j}, v_{k}] + \halfi \eijk \mathcal{D}_j C_k$ \end{tabular} 
& \begin{tabular}{l} $-\halfi [\overline{\phi}, C_i] - \halfi [\psi_i, v] + \halfi [ \psi_4, v_{i}] - \half [A_4, \chi_{i}]$ \\
$- \halfi \eijk [ \psi_j, v_{k}] + \halfi \eijk \mathcal{D}_j \chi_{k} + \halfi \mathcal{D}_i \chi $ \end{tabular} \\
& & \\
$L$  
& \begin{tabular}{l} $\halfi [\phi, \chi] + \halfi [\lambda, v] + \half [A_4, C_4] - \halfi [ \lambda_{j}, v_{j}]$ \\
$+ \halfi \mathcal{D}_j C_j$ \end{tabular}
& \begin{tabular}{l} $-\halfi [\overline{\phi}, C_4] - \halfi [\psi_4, v] + \half [A_4, \chi] - \halfi [\psi_j, v_{j}] $ \\ 
$+ \halfi \mathcal{D}_j \chi_{j}$ \end{tabular} \\ \hline
\end{tabular}\\

\subsection{Supertransformation in $N=4, D=3$} 

{\bf A-model}\\

\begin{tabular}{c|c||l|l |}
 & gh\# & $s$ & $\overline{s}$ \\ \hline
$\phi$ & $2$ & $0$ & $\chi$ \\
$N$ & $0$ & $-\halfi \chi$ & $-\halfi \psi$ \\
$\overline{\phi}$ & $-2$ & $\psi$ & $0$ \\ \hline
$\psi_\mu$ & $1$ & $-i \mathcal{D}_\mu \phi$ & $H_\mu - \mathcal{D}_\mu N - \half \epsilon_{\mu \rho\sigma} F_{\rho \sigma}$ \\
$\chi_\mu$ & $-1$ & $-H_\mu - \mathcal{D}_\mu N + \half \epsilon_{\mu \rho \sigma} F_{\rho \sigma}$ 
	& $-i \mathcal{D}_\mu \overline{\phi}$ \\ \hline
$\chi$ & $1$ & $-[N, \phi] $ & $-\halfi [\overline{\phi}, \phi] $ \\ 
$\psi$ & $-1$ & $- \halfi [\phi, \overline{\phi}]$ & $ -[N, \overline{\phi}] $ \\ \hline
$A_\mu$ & $0$ & $-\halfi \psi_\mu$ & $-\halfi \chi_\mu$ \\ 
$H_\mu$ & $0$ 
	& \begin{tabular}{l} $\halfi \mathcal{D}_\mu \chi - \halfi \epsilon_{\mu \rho \sigma} \mathcal{D}_\rho \psi_\sigma $ \\
	\hspace*{10mm} $+ \halfi [\phi, \chi_\mu] + \half [\psi_\mu, N] $ \end{tabular}
	& \begin{tabular}{l} $-\halfi \mathcal{D}_\mu \psi - \halfi \epsilon_{\mu \rho \sigma} \mathcal{D}_\rho \chi_\sigma $ \\
	\hspace*{10mm} $- \halfi [\overline{\phi}, \psi_\mu] -\half [\chi_\mu, N]$ \end{tabular} \\ \hline
\end{tabular}\\

\newpage
\noindent
{\bf B-model}\\ 

\begin{tabular}{c|c||l|l|}
 & gh\# & $s$ & $\overline{s}$ \\ \hline
$V_\mu$ & $0$ & $\lambda_\mu$ & $\tilde{\lambda}_\mu$ \\
$A_\mu$ & $0$ & $-i \lambda_\mu$ & $-i \tilde{\lambda}_\mu$ \\ \hline
$\lambda$ & $1$ & $-G$ & $-\halfi \mathcal{D}_\mu V_\mu + K$ \\
$\tilde{\lambda}$ & $-1$ & $-\halfi \mathcal{D}_\mu V_\mu - K$ & $-\tilde{G}$ \\ \hline
$\lambda_\mu$ & $1$ & $0$ & $-\qrt \epsilon_{\mn \rho} \maruk{F_{\nu \rho} + i [V_\nu, V_\rho] + 2i \mathcal{D}_\nu V_\rho}$ \\
$\tilde{\lambda}_\mu$ & $-1$ & $\qrt \epsilon_{\mn \rho} \maruk{F_{\nu \rho} + i [V_\nu, V_\rho] + 2i \mathcal{D}_\nu V_\rho}$ & $0$ \\ \hline
$G$ & $2$ & $0$ & $-i \mathcal{D}_\mu \lambda_\mu -i [V_\mu, \lambda_\mu]$ \\ 
$\tilde{G}$ & $-2$ & $-i \mathcal{D}_\mu \tilde{\lambda}_\mu -i [V_\mu, \tilde{\lambda}_\mu]$ & $0$ \\
$K$ & $0$ & $-\halfi \mathcal{D}_\mu \lambda_\mu -\halfi [V_\mu, \lambda_\mu]$ 
	& $\halfi \mathcal{D}_\mu \tilde{\lambda}_\mu +\halfi [V_\mu, \tilde{\lambda}_\mu]$ \\ \hline
\end{tabular}\\

\subsection{Supertransformation in $N=4, D=2$ A-model SYM} 

{\bf A-model}\\ 

\begin{tabular}{c|c||l|l|}
 & gh\# & $s$ ($+1$) & $\overline{s}$ ($-1$) \\ \hline
$\phi$ & $0$ & $0$ & $\chi$ \\
$N$ & $0$ & $-\halfi \chi$ & $-\halfi \psi$ \\
$\overline{\phi}$ & $0$ & $\psi$ & $0$ \\
$A_3$ & $0$ & $-\halfi \psi_3$ & $-\halfi \chi_3$ \\ \hline
$\psi_\mu$ & $1$ & $-i \mathcal{D}_\mu \phi$ & $H_\mu - \mathcal{D}_\mu N - \emn \mathcal{D}_\nu A_3$ \\
$\chi_\mu$ & $-1$ & $-H_\mu -\mathcal{D}_\mu N + \emn \mathcal{D}_\nu A_3$ 
	& $-i \mathcal{D}_\mu \overline{\phi}$ \\ \hline
$\psi_3$ & $1$ & $-[A_3, \phi]$ & $H_3 + i[A_3, N] - \half \emn F_{\mn}$ \\
$\chi_3$ & $-1$ & $-H_3 + i[A_3, N] + \half \emn F_{\mn}$ & $-[A_3, \overline{\phi}]$ \\ \hline
$\chi$ & $1$ & $-[N, \phi]$ & $-\halfi [\overline{\phi}, \phi]$ \\
$\psi$ & $-1$ & $- \halfi [\phi, \overline{\phi}]$ & $-[N, \overline{\phi}]$ \\ \hline
$A_\mu$ & $0$ & $-\halfi \psi_\mu$ & $-\halfi \chi_\mu$ \\
& & & \\
$H_\mu$ & $0$ & \begin{tabular}{l} $\halfi \mathcal{D}_\mu \chi - \halfi \emn \mathcal{D}_\nu \psi_3$ \\
	$+ \half \emn [A_3, \psi_\nu] + \halfi [\phi, \chi_\mu] + \half [\psi_\mu, N]$ \end{tabular} 
	& \begin{tabular}{l} $-\halfi \mathcal{D}_\mu \psi - \halfi \emn \mathcal{D}_\nu \chi_3$ \\
	$+ \half \emn [A_3,\chi_\nu] - \halfi [\overline{\phi}, \psi_\mu] - \half [\chi_\mu, N] $ \end{tabular}\\
& & & \\
$H_3$ & $0$ & \begin{tabular}{l} $\half [A_3, \chi] - \halfi \emn \mathcal{D}_\mu \psi_\nu$ \\
	$+ \halfi [\phi, \chi_3] + \half [\psi_3, N]$ \end{tabular} 
	& \begin{tabular}{l} $-\half [A_3, \psi] - \halfi \emn \mathcal{D}_\mu \chi_\nu$ \\
	$-\halfi [\overline{\phi}, \psi_3] - \half [\chi_3, N] $ \end{tabular} \\ \hline
\end{tabular}\\

\vspace*{10mm}

\noindent
\begin{tabular}{c|c||l|l|}
 & gh\# & $s_3$ ($-1$) & $\overline{s}_3$ ($+1$) \\ \hline
$\phi$ & $2$ & $\psi_3$ & $0$ \\
$N$ & $0$ & $\halfi \chi_3$ & $\halfi \psi_3$ \\
$\overline{\phi}$ & $-2$ & $0$ & $\chi_3$ \\ 
$A_3$ & $0$ 
	& $-\halfi \psi $ & $-\halfi \chi $ \\ \hline
$\psi_\mu$ & $1$ 
	& $\epsilon_{\mu \rho} H_\rho - \mathcal{D}_\mu A_3 + \epsilon_{\mu \rho} \mathcal{D}_\rho N $ 
	& $-i \epsilon_{\mu \rho} \mathcal{D}_\rho \phi $ \\
$\chi_\mu$ & $-1$
	& $i \epsilon_{\mu \rho} \mathcal{D}_\rho \overline{\phi} $ 
	& $\epsilon_{\mu \rho} H_\rho - \mathcal{D}_\mu A_3 - \epsilon_{\mu \rho} \mathcal{D}_\rho N $ \\ \hline
$\psi_3$ & $1$ & $ -\halfi [\overline{\phi},\phi] $ & $ [N, \phi]$ \\
$\chi_3$ & $-1$ & $ [N, \overline{\phi}]$ & $- \halfi  [\phi, \overline{\phi}]$ \\ \hline
$\chi$ & $1$ & $-H_3 -i [A_3, N] + \half \epsilon_{\rho \sigma} F_{\rho \sigma}$ & $- [A_3, \phi] $ \\
$\psi$ & $-1$ & $- [A_3, \overline{\phi}] $ & $H_3 -i [ A_3, N] - \half \epsilon_{\rho \sigma} F_{\rho \sigma}$ \\ \hline
$A_\mu$ & $0$ 
	& $-\halfi \epsilon_{\mu \rho} \chi_\rho $ & $ \halfi \epsilon_{\mu \rho} \psi_\rho$ \\
 & & & \\
$H_\mu$ & $0$ 
	& \begin{tabular}{l} 
	$\halfi (\epsilon_{ \mu \rho} \mathcal{D}_\rho \psi -i [A_3, \chi_\mu] - \mathcal{D}_\mu \chi_3) $ \\
	\hspace*{10mm} $-\half \epsilon_{\mu \rho} [N, \chi_\rho] + \halfi \epsilon_{\mu \rho} [\overline{\phi}, \psi_\rho] $ \end{tabular}
	& \begin{tabular}{l}
	$\halfi (\epsilon_{\mu \rho} \mathcal{D}_\rho \chi + i [ A_3, \psi_\mu] + \mathcal{D}_\mu \psi_3) $ \\
	\hspace*{10mm} $-\half \epsilon_{\mu \rho} [N, \psi_\rho] + \halfi \epsilon_{\mu \rho} [\phi, \chi_\rho] $ \end{tabular} \\ 
 & & & \\
$H_3$ & $0$ 
	& $\halfi \mathcal{D}_\rho \chi_\rho + \half [A_3, \chi_3] + \halfi [\overline{\phi}, \chi] - \half [\psi, N] $ 
	& $ -\halfi \mathcal{D}_\rho \psi_\rho - \half [A_3, \psi_3] - \halfi [\phi, \psi] + \half [\chi, N] $ \\ \hline
\end{tabular}

\subsection{Supertransformation in $N=4, D=2$ B-model SYM}
\begin{tabular}{c|c||l|l|}
 & gh\# & $s$ & $\overline{s}$ \\ \hline
$V_\mu$ & $0$ & $\lambda_\mu$ & $\tilde{\lambda}_\mu$ \\
$V_3$ & $0$ & $\lambda_3$ & $\tilde{\lambda}_3$ \\
$A_\mu$ & $0$ & $-i \lambda_\mu$ & $-i \tilde{\lambda}_\mu$ \\
$A_3$ & $0$ & $-i \lambda_3$ & $-i \tilde{\lambda}_3$ \\ \hline 
$\lambda$ & $1$ & $-G$ & $-\halfi \mathcal{D}_\mu V_\mu - \half [A_3, V_3] + K$ \\
$\tilde{\lambda}$ & $-1$ & $-\halfi \mathcal{D}_\mu V_\mu - \half [A_3, V_3] -K$ & $-\tilde{G}$ \\ \hline
$\lambda_\mu$ & $1$ & $0$ 
	& \begin{tabular}{l} $- \emn \left( \half \mathcal{D}_\nu A_3 + \halfi [V_\nu, V_3] \right. $ \\
	\hspace*{15mm} $\left. + \halfi \mathcal{D}_\nu V_3 - \half [A_3, V_\nu] \right) $ \end{tabular}\\
$\tilde{\lambda}_\mu$ & $-1$ 
	& \begin{tabular}{l} $\emn \left( \half \mathcal{D}_\nu A_3 + \halfi [V_\nu, V_3] \right. $ \\
	\hspace*{15mm} $\left. +\halfi \mathcal{D}_\nu V_3 -\half [A_3, V_\nu] \right) $ \end{tabular} & $0$ \\ \hline
$\lambda_3$ & $1$ & $0$ & $-\qrt \emn \maruk{F_{\mn} + i[V_\mu, V_\nu] + 2i \mathcal{D}_\mu V_\nu}$  \\
$\tilde{\lambda}_3$ & $-1$ & $\qrt \emn \maruk{F_{\mn} + i[V_\mu, V_\nu] + 2i \mathcal{D}_\mu V_\nu}$ & $0$ \\ \hline
$G$ & $2$ & $0$ & 
	\begin{tabular}{l} $-i \mathcal{D}_\mu \lambda_\mu -[A_3,\lambda_3] $ \\ 
	\hspace*{10mm} $-i[V_\mu, \lambda_\mu] -i[V_3,\lambda_3]$ \end{tabular} \\
$\tilde{G}$ & $-2$ & 
	\begin{tabular}{l} $-i \mathcal{D}_\mu \tilde{\lambda}_\mu - [A_3, \tilde{\lambda}_3] $ \\
	\hspace*{10mm} $-i [V_\mu, \tilde{\lambda}_\mu] -i[V_3, \tilde{\lambda}_3]$ \end{tabular} & $0$ \\
$K$ & $0$ & 
	\begin{tabular}{l} $-\halfi \mathcal{D}_\mu \lambda_\mu - \half[A_3, \lambda_3] $ \\ 
	\hspace*{10mm} $- \halfi [V_\mu,\lambda_\mu] -\halfi [V_3, \lambda_3]$ \end{tabular}
& \begin{tabular}{l} $\halfi \mathcal{D}_\mu \tilde{\lambda}_\mu + \half [A_3, \tilde{\lambda}_3] $ \\
	\hspace*{10mm} $+\halfi[V_\mu, \tilde{\lambda}_\mu ]+ \halfi [V_3, \tilde{\lambda}_3]$ \end{tabular} \\ \hline
\end{tabular}\\

\vspace*{10mm}

\noindent
\begin{tabular}{c|c||l|l|}
 & gh\# & $s_3$ & $\overline{s}_3$ \\ \hline
$V_\mu$ & $0$ & $- \epsilon_{\mu \rho} \tilde{\lambda}_\rho$ & $\epsilon_{\mu \rho} \lambda_\rho$ \\
$V_3$ & $0$ & $\tilde{\lambda}$ & $\lambda$ \\
$A_\mu$ & $0$ & $- i\epsilon_{\mu \rho} \tilde{\lambda}_\rho$ & $i \epsilon_{\mu \rho} \lambda_\rho$ \\
$A_3$ & $0$ & $-i \tilde{\lambda}$ & $-i \lambda$ \\ \hline
$\lambda$ & $1$ & $\qrt \epsilon_{\rho \sigma} \maruk{ F_{\rho \sigma } + i[V_\rho, V_\sigma] -2i \mathcal{D}_\rho V_\sigma }$ 
	 & $0$ \\
$\tilde{\lambda}$ & $-1$ & $0$ 
	& $-\qrt \epsilon_{\rho \sigma} \maruk{ F_{\rho \sigma } + i[V_\rho, V_\sigma] -2i \mathcal{D}_\rho V_\sigma }$ \\ \hline
$\lambda_\mu$ & $1$ 
	& \begin{tabular}{l} $-\half \mathcal{D}_\mu A_3 - \halfi [V_3, V_\mu]$ \\ 
	\hspace*{10mm} $- \half [A_3, V_\mu] - \halfi \mathcal{D}_\mu V_3$ \end{tabular} & $0$ \\
$\tilde{\lambda}_\mu$ & $-1$ & $0$ 
	& \begin{tabular}{l} $- \half \mathcal{D}_\mu A_3 - \halfi [V_3, V_\mu] $ \\ 
	\hspace*{10mm} $- \half [A_3,V_\mu] -\halfi \mathcal{D}_\mu V_3$ \end{tabular} \\ \hline
$\lambda_3$ & $1$ 
	& $- \half [A_3, V_3] + \halfi \mathcal{D}_\rho V_\rho +  K $  & $G$ \\
$\tilde{\lambda}_3$ & $-1$ & $\tilde{G}$ 
	& $- \half [A_3, V_3] + \halfi \mathcal{D}_\rho V_\rho - K $ \\ \hline
$G$ & $2$ 
	& \begin{tabular}{l} $ [A_3, \lambda] - i \epsilon_{\rho \sigma} \mathcal{D}_\rho \lambda_\sigma $ \\ 
	\hspace*{10mm} $+ i [V_3, \lambda] + i \epsilon_{\rho \sigma} [V_\rho, \lambda_\sigma]$ \end{tabular} & $0$ \\ 
$\tilde{G}$ & $-2$ & $0$ 
	& \begin{tabular}{l} $[A_3, \tilde{\lambda}] + i \epsilon_{\rho \sigma} \mathcal{D}_\rho \tilde{\lambda}_\sigma $ \\
	\hspace*{10mm} $+ i [V_3, \tilde{ \lambda} ] - i \epsilon_{\rho \sigma} [V_\rho, \tilde{\lambda}_\sigma]$ \end{tabular} \\ 
$K$ & $0$ 
	& \begin{tabular}{l} $\half [A_3, \tilde{\lambda}] + \halfi \epsilon_{\rho \sigma} \mathcal{D}_\rho \tilde{\lambda}_\sigma $ \\
	\hspace*{10mm} $+ \halfi [V_3, \tilde{\lambda}] - \halfi \epsilon_{\rho \sigma} [V_\rho, \tilde{\lambda}_\sigma]$ \end{tabular}
	& \begin{tabular}{l} $-\half [A_3, \lambda] + \halfi \epsilon_{\rho \sigma} \mathcal{D}_\rho \lambda_\sigma$ \\
	\hspace*{10mm} $- \halfi [V_3, \lambda] - \halfi \epsilon_{\rho \sigma} [V_\rho, \lambda_\sigma] $ \end{tabular} \\ \hline
\end{tabular} \\

\end{document}